# An Approach for the Automation of IaaS Cloud Upgrade

Mina Nabi
Ericsson Inc.
Ottawa, Canada
mina.nabi@ericsson.com

Ferhat Khendek
Electrical and Computer Engineering
Concordia University
Montreal, Canada
ferhat.khendek@concordia.ca

Maria Toeroe
Ericsson Inc.
Montreal, Canada
Maria.Toeroe@ericsson.com

***Abstract*— An Infrastructure as a Service (IaaS) cloud provider is committed to each tenant by a service level agreement (SLA) which indicates the terms of commitment, e.g. the level of availability of the IaaS cloud service. The different resources providing this IaaS cloud service may need to be upgraded several times throughout their life-cycle; and these upgrades may affect the service delivered by the IaaS layer. This may violate the SLAs towards the tenants and result in penalty as they impact the tenants' services relying on the IaaS. Therefore, it is important to handle upgrades properly with respect to the SLAs.**

**The upgrade of IaaS cloud systems inherits all the challenges of clustered systems and faces other, cloud specific challenges, such as size and dynamicity due to elasticity. In this paper, we propose a novel approach to automatically upgrade an IaaS cloud system under SLA constraints such as availability and elasticity. In this approach, the upgrade methods and actions appropriate for each upgrade request are identified, scheduled, and applied automatically in an iterative manner based on: the vendors' descriptions of the infrastructure components, the tenant SLAs, and the status of the system. The proposed approach allows new upgrade requests during ongoing upgrades, which makes it suitable for continuous delivery. In addition, it also handles failures of upgrade actions through localized retry and undo operations automatically.**

***Keywords*— High Availability; Cloud; IaaS; Upgrade; Scaling;**

## I. INTRODUCTION

In the Infrastructure as a Service (IaaS) cloud model, there are three types of resources: physical resources, virtualization facility resources, and virtual resources [1]. During their life-cycle these resources undergo potentially several upgrades, which impact the services they provide, hence, may induce service outage. An IaaS cloud provider is committed to a tenant by a *service level agreement (SLA)* which indicates the terms of the commitment [2], e.g. the percentage of time the service must be accessible (referred to as *availability* requirement). To avoid penalties due to SLA violation such as service outages, the upgrade of an IaaS cloud system has to be carried out carefully with minimal impact on the provided service.

There are several challenges of maintaining availability during IaaS upgrades [1]. Some of these challenges are similar to traditional clustered systems, while others are specific to the cloud. As in clustered systems, handling of dependencies is important to prevent service outages during upgrades. Within the IaaS layer itself resources may depend on each other; and, of course, all other cloud layers rely and depend on the resources of the IaaS layer. These dependencies require compatibility. During the upgrade process, incompatibilities may arise and break dependencies that do not exist in the current/source or in the desired configurations. Therefore, the upgrade methods and their actions need to be selected and orchestrated carefully with respect to the dependencies and their potential incompatibilities. Specific upgrade methods handling incompatibilities may also require additional resources. Considering the scale of an IaaS cloud system, minimizing the amount of such additional resources is also a significant challenge in comparison to clustered systems. As for clustered systems upgrade actions performed on the cloud resources may fail. These failures, henceforth referred to as *upgrade failures*, have to be handled adequately to guarantee the consistency of the system configuration.

The dynamicity of cloud systems introduces additional challenges to upgrades [3]. Cloud systems adapt to workload changes by provisioning and de-provisioning resources automatically according to the workload variations. This mechanism is referred to as autoscaling [4][5] or elasticity [6]. This dynamicity poses a challenge of satisfying SLAs during upgrades. Indeed, the autoscaling feature may interfere with the upgrade process in different ways. The service capacity of the system decreases during upgrade when resources are taken out of service for the upgrade. In the meantime, the system might need to scale out in response to workload increase creating contention. Furthermore, the autoscaling may undo or hinder the upgrade process when scaling in releases newly upgraded resources (e.g. VMs), or when scaling out uses the old (i.e. not yet upgraded) version of the resources. Therefore to avoid such issues, autoscaling is generally disabled during upgrades as reported in [7][8].

Different upgrade methods (e.g. rolling upgrade [9], split mode [10] and delayed switch [11]) have been proposed for maintaining *high availability (HA)*, i.e. the availability of services at least 99.999% of the time, during the upgrade of clustered systems. However, none of these methods addresses all of the challenges of upgrades in the cloud environment. For instance, Windows Azure Storage uses rolling upgrade to partition the system into subsystems and upgrade them one at a time [12]. The rolling upgrade method may introduce mixed-version inconsistencies in case of incompatibility between the different versions of redundant resources [13][14]. Other solutions propose the parallel universe [13][15][16] method to address such incompatibilities. In this case, an entire new system is created with the new configuration, while the old system continues serving. The two systems exist in parallel until the new system can take over serving. As a result, applying this method can be very costly.

Equally important, there is no method proposed to automate the entire IaaS upgrade process spanning from selecting the appropriate upgrade methods to orchestrating the upgrade process while avoiding or at least limiting service outages during the upgrade. Automation is crucial due to the size of cloud deployments and for supporting zero-touch operations.

In a previous work [3] we proposed a method for upgrading the IaaS compute resources while maintaining the availability of compute services according to some SLA parameters. This method was applicable to compute hosts upgrades such as the upgrade of the hypervisor, the host OS, or the physical host. The method in [3] does not handle the upgrade of other kinds of IaaS resources, including network or storage, as the dependencies between those types of resources have not been taken into account. For example, the IaaS is responsible of providing an image storage service to the IaaS cloud throughout its life-cycle, which includes upgrades. If the storage resources are to be upgraded, this dependency becomes a particular challenge as it supports the VM related operations (e.g. live migration) used by the upgrade process itself. Since the scope of the approach in [3] was limited to IaaS compute resources, handling of other kinds of resources with respect to their dependencies was out of scope. In addition, upgrade failures were not handled in [3].

In this paper, we remove all these limitations and propose a novel approach for managing the upgrade of different kinds of resources of an IaaS cloud system while also considering the SLA constraints for availability and elasticity. In this approach, the service disruption caused by the upgrade of the IaaS cloud system is minimized by orchestrating the entire upgrade process. Each upgrade request is applied in accordance with the SLA parameters considered (i.e. availability and elasticity), the current status of the system, and the information provided by vendor in the form of an *infrastructure component description*, which is a file accompanying a product delivered by a vendor indicating the infrastructure component(s) delivered as part of the product and all the information necessary for deploying these infrastructure component(s) in the IaaS layer.

The proposed approach identifies the upgrade actions necessary for each IaaS resource and the upgrade method appropriate for applying those actions taking into account different dependencies. Subsequently, it generates upgrade schedules and applies them to the resources in an iterative manner considering availability and elasticity constraints of the SLAs. In case of an upgrade failure, the recovery operations (i.e. retry and undo) are handled automatically to bring the system to a consistent configuration. As a result, and in comparison with existing solutions, our approach addresses all the aforementioned challenges of IaaS cloud upgrades in an integrated manner. Moreover, our approach is capable of handling new upgrade requests even during ongoing upgrades, which makes it suitable for continuous delivery. We prototyped our approach and performed evaluations to compare it with current practices.

The rest of the paper is organized as follows. In Section II, we provide the preliminaries and the definitions of concepts used throughout the paper. In Section III, we present the main principles and the overall framework of our approach before elaborating on the details in Section IV. In Section V, the implementation and the validation of our approach are presented and discussed. In Section VI, we review in more details the related work and we conclude in Section VII.

## II. PRELIMINARIES AND CONCEPT DEFINITIONS

### A. *IaaS Cloud System*

We view an IaaS cloud system as sets of physical hosts providing compute services ($M_{compute}$) and virtual storage ($M_{storage}$), dedicated to network services and to controller services, and sets of other physical resources for networking (e.g. switch, router) and storage (physical storage). Note that $M_{compute}$ and $M_{storage}$ may overlap. The size of any of these sets may change during the upgrade due to failures, administrative operations, and/or the upgrade process itself. In this paper, we assume that all physical hosts in $M_{compute}$ have a capacity to host $K$ VMs of equal size.

The number of tenants may also vary over time. As we apply the changes in an iterative manner, we denote by $N_i$ the number of tenants served by the IaaS cloud at iteration *i*. The $n_{th}$ tenant has a number of VMs, which may vary between a minimum ($min_n$) and a maximum ($max_n$) that the IaaS provider agreed to provide in the respective SLA. The SLA of tenant $n$ also specifies a scaling adjustment $s_n$ value and a cooldown duration $c_n$, which represent the maximum size of the adjustment in terms of VMs in one scaling operation to be satisfied by the IaaS provider and the minimum amount of time between two subsequent scaling operations. These parameters (i.e. $min_n$, $max_n$, $s_n$, $c_n$) define the SLA elasticity constraints imposed by the $n_{th}$ tenant.

The definitions of the parameters used in this paper are provided in Table A in the appendix.

We assume that the IaaS cloud system is configured as a highly available system, which includes a service for managing VM availability such as the Senlin [17] clustering service for OpenStack. The availability of the applications deployed in the VMs is managed by an availability management solution such as the Availability Management Framework (AMF) [18], as proposed in [19] for instance. The requirements of the application level redundancy are expressed towards the IaaS cloud as VM placement constraints (i.e. as anti-affinity groups), which need to be respected during the upgrade to satisfy the availability SLA constraints. VMs of a tenant may form several anti-affinity placement groups. This means not only that VMs of the same group should be placed on different physical hosts, but also that at most a specific number of VMs of a tenant can be impacted at a time. As availability constraints we consider these VM placement constraints.

Fig. 1 shows an illustrative example of a system with 15 hosts. Nine of these hosts participate in the creation of a VMware Virtual Storage Area Network (VSAN) – the storage supporting VM operations in the system ($|M_{storage}|=9$), while 10 of the hosts provide compute services ($|M_{compute}|=10$). Thus, host 6 through host 9 belong to both sets. In this example we assume that each host in $M_{compute}$ has a capacity to serve two VMs ($K=2$). In addition to these resources, there are dedicated network resources: switches (denoted by SW) and routers (denoted by R) shown at the bottom of the figure. We assume four tenants with their respective elasticity constraints. Note that the controller hosts are not shown in Fig. 1.

The IaaS resources may have different dependencies. The dependencies considered by our approach have been defined in [1], except for the aggregation dependency and the VM supporting storage/controller dependency, which we define here. The *aggregation dependency* reflects the fact that multiple resources of the same kind may be used to create a new resource of a different kind, i.e. the aggregate resource, which can function as long as (i.e. it depends on) a minimum number of constituent resources are functioning (i.e. sponsoring). The aggregate resource will experience an outage whenever the number of available constituent resources drops below the minimum required. This is the case, for instance, with a virtual shared storage built using a cluster of physical disks, i.e. the VSAN of our example. The *VM supporting storage/controller dependency* indicates a dependency of a compute host resource on a storage or controller resource for VM operations. For instance, the storage infrastructure resource provides the storage service, which supports the VM operations running on the dependent compute hosts and which are used during upgrades.

### B. Infrastructure Component and Resource Upgrade Catalog

We define an *infrastructure component* as a piece of software, firmware, or hardware delivered by a vendor as part of a product. The product itself can be simple with a single component (e.g. ESXi hypervisor) or a compound product consisting of different components (e.g. Ceph storage with different components). Once a product is fully installed in the IaaS system, it can be instantiated as an IaaS resource (e.g. ESXi hypervisor, Ceph storage). We assume each infrastructure component, delivered as part of a product, is accompanied with a machine-readable file referred to as the infrastructure component description. This file describes at least the component's service capabilities, configuration constraints, hardware management capabilities, delivering software/firmware bundle with their installation/removal and activation/deactivation scripts/commands, estimated time required for their installation/removal, and hardware/software dependencies. E.g. the information expected is similar to the content of the OVF descriptor [20] used for describing virtual appliances with some extensions such as the times required for installation/removal of the components.

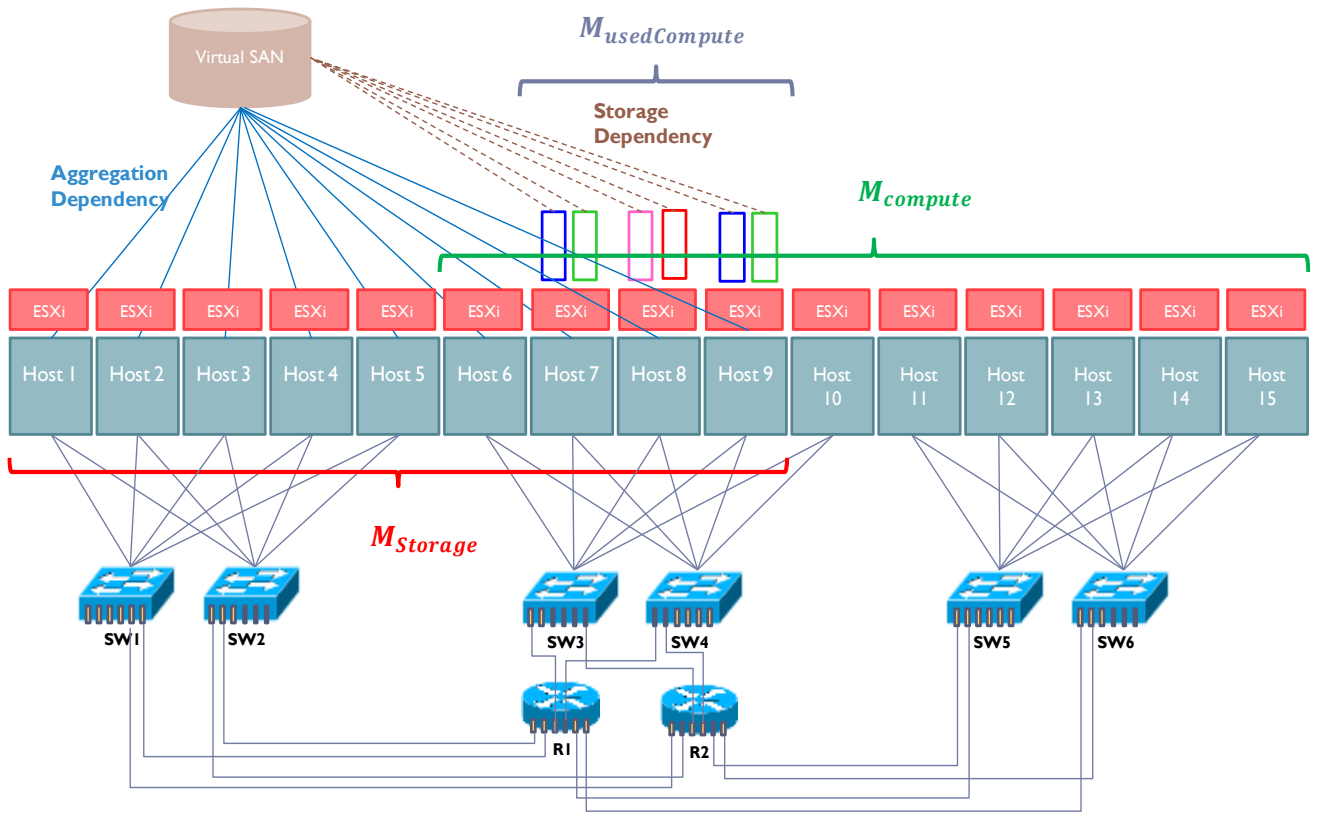

Fig. 1. Illustrative example

To automate the upgrade of IaaS cloud system, all the information included in the infrastructure component descriptions needs to be accessible at any time. For this purpose, we define and use a *Resource Upgrade Catalog*. This catalog includes all the infrastructure component descriptions provided by the vendors for all the components already deployed in the system and those to be added to the system. In our illustrative example, the resource upgrade catalog initially includes, among others, the infrastructure component descriptions for VSAN and ESXi. If we would like to upgrade it to Ceph, the infrastructure component descriptions of Ceph also need to be added to the catalog. Using these infrastructure component descriptions, the scripts for upgrading the virtual shared storage from VSAN to Ceph can be derived. The same applies also for upgrading from Ceph to VSAN should an undo become necessary.

### C. Changes, Actions, Operations and Units

A *change* in the IaaS cloud system configuration can be the addition, removal, or upgrade of an infrastructure component of some resources, some resources themselves, or a dependency between two resources or their sets. To deploy a change in the IaaS cloud system configuration, one or more upgrade actions may need to be executed. We define an *upgrade action* as an atomic action that can be executed by a configuration management tool (e.g. Ansible [21]) on a resource (e.g. a command for installing ESXi on a host), or performed by an administrator on a resource (e.g. hardware swap). An upgrade action is always associated with one or more *undo actions* which are upgrade actions that revert the effect of the upgrade action applied on a resource and which they are associated with. We use the expression *upgrade operation* as a shorthand for an ordered list of upgrade actions. Thus, the upgrade actions deploying a change can be referred to as an upgrade operation. We use similarly the expression *undo operation*.

We define the concept of *undo unit*. It consists of a group of resources on which an upgrade operation has to be applied successfully all together; otherwise, the respective undo operation is triggered.

We also define an *upgrade unit* as a group of resources that need to be upgraded using an appropriate upgrade method, for example, for handling incompatibilities.

### D. Upgrade Requests

The system administrator indicates the desired changes of IaaS resources by specifying upgrade requests. Each *upgrade request* is a collection of change sets, i.e. a set of change sets. Each *change set* in the collection specifies one or more tightly coupled changes that should either succeed or fail together to maintain the consistency of the system configuration. That is, in terms of resources a change set identifies an undo unit. Different change sets, on the other hand, are independent of each other, and the failure of one does not impact the success of other change sets. Whenever the administrator issues an upgrade request with a change set referring to a product the resource upgrade catalog is checked. If the product and its accompanying infrastructure component descriptions are not available yet, they need to be added to the catalog first. Since the administrator may not be aware of all the software and hardware dependencies of a product and may not specify all the details necessary as changes in the change set, some change sets may be incomplete or not detailed enough. To satisfy all hardware and/or software dependencies and determine the details necessary for deployment, we check each change set with respect to the infrastructure component description(s) provided by the vendor(s) and the system configuration, and we identify the detailed as well as any missing changes. These detailed/missing changes are added to the change set as complementary changes.

For each requested change we also derive from the infrastructure component description(s) the necessary upgrade actions, that is, for software components the scripts used to install/remove them, while for hardware components the scripts used for their management.

Considering our illustrative example of Fig. 1, an administrator may want to issue an upgrade request with two changes: (1) to upgrade the virtual shared storage from VSAN to Ceph; and (2) to upgrade the networking infrastructure from IPv4 to IPv6. These changes are independent of each other, therefore the administrator separates them into two change sets. For each set, the complementary changes will be inferred automatically from the infrastructure component descriptions considering the system configuration of Fig. 1. The first change in details requires detaching the hosts from the VSAN cluster, upgrading their hypervisors from ESXi to a hypervisor supported by Ceph (e.g. QEMU or Xen), and configuring the Ceph components (e.g. OSD, monitoring, and client daemons) on the hosts. The second change implies the upgrade of all routers, switches and hosts to IPv6. These changes are added as complementary changes to the first and second change sets given by the administrator in the upgrade request.

The administrator can control the upgrade execution by specifying four additional parameters in the upgrade request. A max-completion-period and a max-retry threshold can be specified for each change set, to ensure the completion of the upgrade process. The *max-completion-period* indicates the maximum time that all the changes of the set may take to complete. The *max-retry* threshold indicates the maximum number of permitted upgrade attempts on each resource to which a change in that change set is applied. In addition, an undo-threshold and an undo-version can be specified for each change. Each change in a change set is applicable to a set of resources. The *undo-threshold* indicates the minimum required number of resources in this set of resources that should be operational after applying the requested change. The *undo-version* parameter specifies the desired version for the undo operation. The default undo-version is the version at which a resource is at the moment the change is applied. For the new upgrade requests, which are issued during ongoing upgrades, the undo-version may not be deterministic. Therefore, the default undo-version can be overridden by explicitly specifying the undo-version in the change request.

We keep track of the upgrade requests using an *upgrade request model*. It maintains the information regarding the execution status (i.e. new, scheduled, completed, or failed) of the change sets and of the changes within each set, and it is updated as the upgrade progresses. Whenever a new upgrade request is issued, its change sets (including any complementary changes) are added to the upgrade request model. For each change in a change set, this model includes the information of target resources, as well as the source, target and undo-versions of each resource.

### E. Resource Graph

To coordinate the upgrade process, it is necessary to keep track of the configuration of the system as well as the status of the ongoing upgrades. For this purpose, we define a *Resource Graph (RG)*, which maintains the state of the upgrade process with respect to IaaS resources and their dependencies. Thus, it is derived from the system configuration and the upgrade requests considering the infrastructure component descriptions, and it is continuously updated during the upgrade process.

The *RG* is a directed graph *(R, D)*, where *R* is the set of vertices representing resources and *D* is the set of edges representing dependencies.

The vertices represent existing or to be added resources in the system. A vertex (resource) is characterized by the following attributes:

- *Resource-id*: the id of the resource. It is created when a new resource is added to the RG and maintained in the system throughout the life-cycle of the resource. For existing resources it is collected from the configuration.
- *Resource-kind*: the kind of the resource (e.g. compute host, switch, router, etc.) according to the infrastructure resource models as described in [1].
- *Modification-type*: it indicates whether the resource is to be upgraded, added, or removed by a requested change, or it remains unchanged. It can have one of the following values: "Upgrade", "Add", "Remove", or "No-change". As the upgrade proceeds, the value of this parameter is updated to reflect the next change to be applied to the resource.
- *Activation-status*: the resource status may be active (i.e. in service) or deactivated (i.e. out of service).

- *Undo-unit-ids*: the set of undo units the resource belongs to. Since there may be several change sets impacting the same resource, each resource may be associated with several undo units.
- *Actions-to-execute:* is an ordered list of execution-levels where each *execution-level* is an ordered list of upgrade actions to be executed on the resource. This allows two levels of coordination of upgrade actions, for the resource itself within an execution-level and between different resources and/or upgrade requests by execution-levels.
- *Number-of-failed-upgrade-attempts*: is the per undo unit counter of the failed upgrade attempts for the resource.
- *Is-isolated:* indicates whether the resource is isolated or not.
- *Is-failed:* indicates whether the resource is failed or not.

*D* is a set of edges, each representing a dependency between resources, either in the source or in the future configuration. The edges can be of different types to capture the different types of dependencies identified in IaaS cloud systems [1]: container/contained dependency, migration dependency, composition dependency, aggregation dependency, communication dependency, storage dependency, controller dependency, VM supporting storage/controller dependency, and peer dependency between resources[1].

An edge $d_{ij}$ denotes a dependency of resource $R_i$ on resource $Rj$, i.e. it is directed from the dependent to the sponsor resource. A symmetrical dependency (i.e. peer dependency) is represented by a pair of edges between two resources, i.e. $d_{ij}$ and $d_{ji}$. Each edge (dependency) has two parameters:

- *Presence*: it indicates whether the dependency exists in the source configuration, in the future configuration, or in both. It can hold the values of “future”, “current”, or “current/future”.
- *Incompatibility-factor*: a “true” value indicates an incompatibility along the dependency, which needs to be managed during the upgrade to avoid outages. Assuming that the source and the desired configurations are consistent within themselves, an incompatibility can only occur along a dependency with the presence value of “current/future”. This parameter is used to identify the upgrade units.

Fig. 2 shows the RG representing the configuration of our illustrative example of Fig. 1 after the upgrade request with the two change sets discussed earlier (i.e. upgrade from VSAN to Ceph and from IPv4 to IPv6) was received. For readability only part of the system configuration and the modification-types requested are shown.

In this RG vertices of R1 to R15 represent the hypervisors running on host1 to host15 represented by vertices R16 to R30. This hosting relation, which is a container/contained dependency is represented by the edges between the vertex pairs e.g. R1 and R16.

As mentioned in Section II.B, a product delivered by a vendor may be a compound product with components to be installed on different IaaS resources. In this example, the existing VSAN virtual shared storage and the desired Ceph virtual shared storage are both compound products. In the source configuration, storage hosts R16 to R24 are aggregated into the virtual shared storage of R46 representing VSAN as shown by the edges between them. In addition, R46 serves as a VM supporting storage to the compute hosts R21 to R30 also reflected as edges.

Since the virtual shared storage is a storage resource supporting VM operations, and the VSAN cannot be upgraded to Ceph in place due to incompatibilities, our approach uses the *Partial Parallel Universe (PPU)* method for this upgrade, which implements parallel universe method locally. This means in terms of detailed changes the addition of the new Ceph virtual shared storage resource followed by the removal of the old VSAN resource. Accordingly, in the RG we use two vertices, one for the old VSAN configuration with modification-type of remove (e.g. R46), and one for the new Ceph configuration with modification-type of add (e.g. R45). To deploy the Ceph product further details are necessary: Its components need to be installed on different IaaS resources, i.e. on storage hosts represented by R16 to R20 to be aggregated into R45, and on compute hosts represented by R21 to R30 to support their VM operations. Since these dependencies exist only in the future Ceph configuration their presence is “future”, while the similar dependencies for VSAN are marked as “current”. These resources are identified based on the requested change, the RG reflecting the current configuration, and the requirements indicated in the Ceph component descriptions.

## III. Overall Principles and Framework for IaaS Cloud Upgrade

In this section we first discuss how we handle the different availability challenges posed by IaaS cloud upgrades, then, we describe the overall framework and the approach.

### A. Handling of Availability Challenges

As mentioned in the introduction, we consider several challenges of maintaining availability during IaaS cloud upgrades: (1) dependency of the application layer on the IaaS layer, (2) resource dependencies within the IaaS cloud, (3) potential incompatibilities along dependencies during the upgrade process, (4) upgrade failures, (5) dynamicity of the cloud environment, and (6) size as well as keeping the amount of additional resources at minimum.

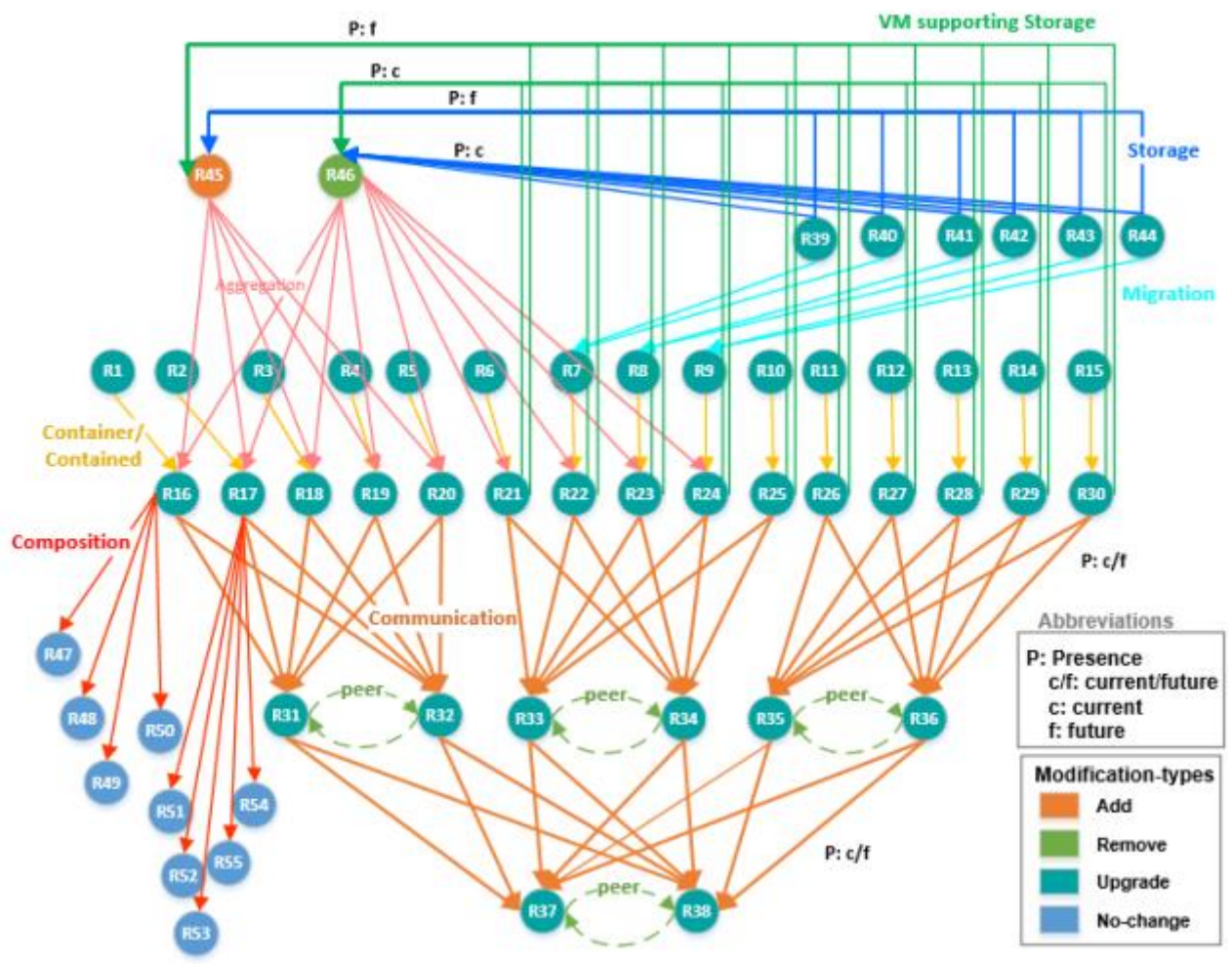

Fig. 2. Partial resource graph for the illustrative example

[1] Note that the aggregation and VM supporting storage/ controller dependencies were not yet identified in [1].

*Dependency of the application layer:* As mentioned earlier we assume that the availability of the applications deployed in the VMs provided by the IaaS layer is managed by an availability management solution and the requirements of the application level redundancy are expressed towards the IaaS cloud as VM placement constraints (i.e. as anti-affinity groups) ensuring that only one VM of an anti-affinity group is impacted by a host failure. The VMs of a tenant may form several anti-affinity groups in worst case placing each VM of the tenant on a different host. Our approach respects these constraints during upgrades, whether VM upgrade, migration or consolidation needs to be performed. That is, it ensures that the VMs of a tenant can always be placed on different physical hosts up to the tenant's maximum number, and at most one VM of a tenant is impacted at a time.

*IaaS resource dependencies:* To maintain the availability of IaaS services we cannot upgrade all the resources at the same time, which means that to avoid breaking any resource dependencies the upgrade of resources must be performed in a specific order depending on the nature of the dependencies. As a solution, we use an *iterative upgrade process* and select at the beginning of each iteration, the resources that can be upgraded in that iteration without violating any dependency. We re-evaluate the situation at the beginning of each subsequent iteration. For the selection of resources upgradeable in the given iteration, first we group into resource groups resources that have to be upgraded at the same time, then using a set of rules – referred to as *elimination rules* – enforcing the dependency constraints, we identify the resource groups that can be upgraded in the current iteration. This results in an initial selection of resource groups referred to as the *initial batch*.

*Potential incompatibilities during upgrade:* Even though the source and the target configurations on their own have no incompatibilities, during the transition of the system from one to the other incompatibilities may occur. That is, during the time of the upgrade, version mismatch may happen along some of the dependencies for some resources. To avoid such incompatibilities, these resources need to be upgraded using an appropriate upgrade method. We identify automatically the resources which have potential incompatibilities that may manifest during their upgrade and group them into upgrade units. To do so we check each dependency against the relevant infrastructure component descriptions provided by the vendors for possible version mismatch. Depending on the type of dependencies with the incompatibilities within the upgrade unit different upgrade methods may be used to handle the situations (such as split mode [10], and others, which are not addressed in details in this paper). We select the most appropriate method and associate it with the upgrade unit. It determines the order of the upgrade of resources within the upgrade unit.

At the resource level each execution-level of a resource is associated with an upgrade unit, which in turn is associated with the selected upgrade method. In each iteration the elimination rules may or may not remove the resource from the initial batch based on the order required by the upgrade method associated with the upgrade unit. For example, if split mode [10] is selected to avoid incompatibilities, the resources of the associated upgrade unit are divided into two partitions which are upgraded one partition at a time. The relevant elimination rules ensure that only one partition is selected at a time, and that the order of deactivation and activation of the partitions is such that it avoids any incompatibilities by having only one version active at any given time.

Whenever a new upgrade request targets the same resources as previous upgrade requests, the new upgrade request may introduce new incompatibilities not considered in the ongoing upgrade. To prevent these incompatibilities to manifest beforehand, new upgrade units are created for the new request with new execution-levels. Thus, the upgrade actions associated with the new upgrade request can only be executed on a resource after finalizing all the upgrade actions of the ongoing upgrade requests.

*Handling of upgrade failures:* In case of upgrade failures, recovery operations are performed to bring the system to a consistent configuration. Since changes in a change set are dependent, there are two main criteria to guarantee a consistent configuration: First, with respect to a given resource all the upgrade actions deploying a change set on that resource must either be applied successfully, or none of them should be applied at all. Second, with respect to different resources impacted by a change set, all the changes of the change set have to be successful without violating their undo thresholds; otherwise, they have to be undone all together.

According to the first criterion, in case an upgrade action of a change set – an upgrade operation – fails, the effects of the already executed upgrade actions of that set need to be reverted. This is referred to as resource level undo and performed immediately. This takes the resource to the version it was before applying the upgrade operation. If this is successful and the retry of the upgrade operation is permitted on the resource, i.e. the applicable max-retry threshold is not reached yet, another attempt can be made to re-execute the upgrade operation. Otherwise, if reverting the upgrade actions was successful (i.e. the previous stable configuration of the resource is reached), but the retry of the operation is not permitted, the resource will be isolated from the system. Hereafter, we refer to such a resource, as an *isolated-only resource*. If reverting the upgrade actions fails, the resource needs to be isolated and marked as *failed*.

According to the second criterion, if the number of isolated-only and failed resources in the set of resources to which a change is applied violates the undo-threshold value, all changes of the change set need to be undone on all applicable resources to preserve system consistency, i.e. required by the second criterion. In this case the undo operation is performed at system level for the change set. It means that all the resources targeted by the change set are checked and are taken to their undo-versions. As mentioned earlier, an undo unit is assigned to each change set and its targeted resources to help identifying the scope of a potential undo operation.

An undo-version different from the original version of the resource needs to be specified explicitly in the upgrade request. Thus, the isolated-only resources may or may not be at their undo-version. If isolated-only resources are at the undo-version, they are released from the isolation. Otherwise an attempt is made to take them to the undo-version. If this is unsuccessful, they are also marked as failed resources.

Since the undo operation itself is set of upgrade actions, thus, it can be performed the same way as other upgrade actions. The only difference is that the upgrade actions used as undo actions are organized into the first execution level of the resources so that they will be executed as soon as possible. An undo operation could be triggered as discussed if the undo-threshold for a set is violated; if all the upgrade actions of the set cannot be finalized within the indicated max-completion-period; or if the administrator explicitly issues an undo operation for a change set that has not been completed yet. Once a change is completed it cannot be undone, instead a new change can be requested.

Whenever a change set is completed successfully the related isolated-only resources are marked as failed (as they were not upgraded successfully) to avoid using them in the target configuration. All failed resources are reported to the administrator, no further automatic action is taken on them.

*Managing the dynamicity of the cloud environment:* To handle the interferences between autoscaling and the upgrade process, we regulate the pace of the upgrade process. To respect the SLA commitments (scaling and availability), in each iteration the current configuration of the system is taken into consideration and only a limited number of resources is taken out of service for upgrade. That is, in each iteration the number of resources necessary for accommodating the current SLA commitments is determined, i.e. the number of resources necessary during that iteration to satisfy potential scaling out requests and to recover from potential failures. These resources need to be reserved and cannot be upgraded without potential violation of availability. So, from the initial batch of resources selected with respect to the dependencies, these reserved resources are also eliminated to select the *final batch* for the iteration, i.e. the resources to be upgraded in this iteration.

The upgrade process starts/resumes if and only if we can take out at least one resource (i.e. the final batch is not empty) and upgrade them without violating the availability and elasticity constraints should there be a resources failure, or should we receive valid scaling requests. If no resource can be taken out, the upgrade process is suspended until there is enough resources, for example, freed up through scaling in requests.

*The challenge of size and minimizing the amount of additional resources:* As we have seen with respect to the dynamicity of the cloud environment, we control the pace of the upgrade process according to the availability of resources for upgrade. On one hand, we try to upgrade in each iteration as many resources as possible in parallel to shorten the upgrade process. On the other hand, we do not consider additional resources, but rather suspend the upgrade process until such resources become available.

However, this may block certain change sets for a long time, therefore, providing additional resources to the system may become temporarily necessary for progressing with the upgrade. The amount may depend on at least the upgrade method, the change set (i.e. the number of resources the upgrade needs to be applied to), and the spare capacity of the system. It may be necessary to add resources to enable the use of certain techniques to maintain service availability and continuity especially in the presence of incompatibilities. As discussed in the introduction, some of the upgrade solutions [13][15][16] use the parallel universe method to avoid incompatibilities, which at the system level is expensive in terms of resources. Our goal is to use in such cases only the minimum additional resources necessary for a given subsystem to keep the cost of the upgrade as low as possible.

To maintain the continuity of the storage or controller services supporting VM operations, when their resources are upgraded and the new and the old versions are incompatible, we propose to use the PPU method. With the PPU method we create a new configuration only for the VM supporting storage and VM supporting controller resources – a subsystem – with their new version while (in parallel) we keep the old version of these resources and their configuration until the new one can take over the supporting role for all the VMs. To transfer the VM supporting role from the old configuration to the new, the physical hosts providing the VM service of the IaaS (i.e. the compute hosts) are also divided into two partitions. The old partition hosts VMs compatible with the old version of the VM supporting storage/controller; it hosts all the VMs initially. The new partition, which is empty initially, hosts the VMs compatible with the new version of the VM supporting storage/controller. As soon as the new versions of the VM supporting storage and controller resources are ready, we migrate the VMs from the old to the new partition potentially in multiple iterations as appropriate for their SLAs. Once all the VMs have been migrated, the old configuration of the VM supporting storage and controller resources can be safely removed.

This procedure means that to guarantee the continuity of the VMs supporting services, the resource requirements for both versions of the configurations of VM supporting storage and controller resources have to be satisfied simultaneously for the time of the upgrade of these resources and until the completion of the VM migrations. If the resource requirements cannot be satisfied using existing resources, additional resources are required at least for the missing amount. This way, we can keep the number of additional resources to a minimum.

### *B. A Framework for IaaS Cloud Upgrade*

For the upgrade of IaaS cloud systems, we propose the upgrade management framework shown in Fig. 3. It includes two main components, the *Upgrade Coordinator* to coordinate the upgrade process according to the principles discussed in Section III.A, and the *Upgrade Engine* to execute the upgrade actions necessary to deploy in the system the requested upgrades.

The upgrade coordinator keeps track of the upgrade requests and decides about the upgrade process in an iterative manner. For each iteration it generates one or more *Runtime Upgrade Schedule(s)*, each of which is a collection of upgrade operations defined through the upgrade actions and the set of resources on which they need to be applied. The upgrade coordinator uses as input the current configuration of the system, the change sets indicated in the upgrade request(s), the infrastructure component descriptions provided by the vendors, and SLAs of the current tenants to generate the schedule for the next iteration. To keep track of the upgrade requests the upgrade coordinator creates an upgrade request model as discussed in Section II.D. Based on the infrastructure component descriptions provided, it infers any complementary changes necessary and identifies the upgrade actions needed to deploy the requested change sets. After each iteration the upgrade coordinator creates an Upgrade Iteration Report.

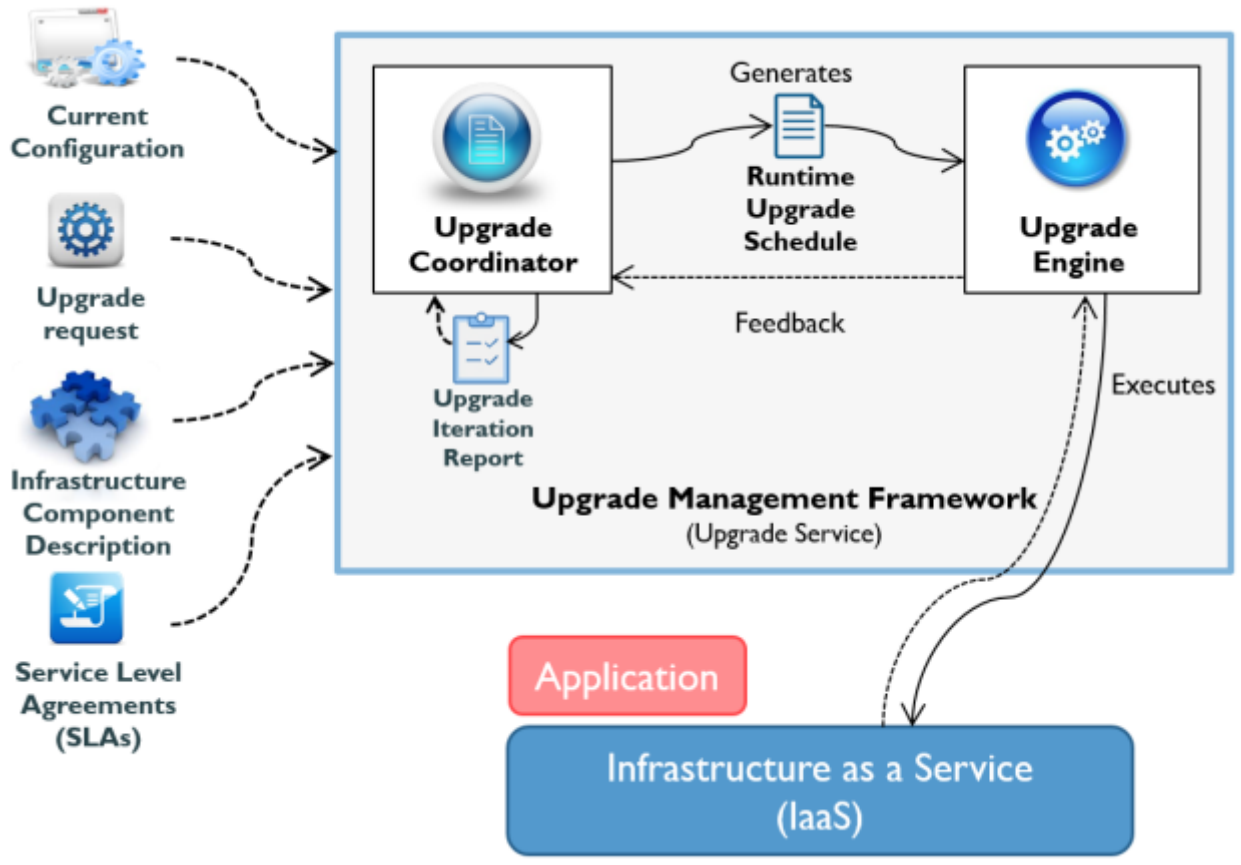

Fig. 3. A framework for IaaS cloud system upgrade

The upgrade engine, an engine capable of running upgrade actions on IaaS resources (e.g. Ansible [21] cloud configuration management tool), executes the upgrade actions specified in the runtime upgrade schedule received from the upgrade coordinator. Note that in case of hardware resources the upgrade engine is limited to management operations and may require administrative assistance for actions such as replacement of a piece of hardware. However, it can bring the resources to the required state and signal when the assistance is necessary and on which piece of hardware.

To maintain service availability, the upgrade of the IaaS cloud system is considered as an iterative process. Fig. 4 illustrates the iterative aspect used in the upgrade coordinator to coordinate the upgrade process. In each iteration, the upgrade coordinator goes through the following four steps to identify the resources that can be upgraded in the current iteration:

- *Step 1- create/update the resource graph*,
- *Step 2- group the IaaS resources for upgrade*,
- *Step 3- select the batch of IaaS resources for upgrade*, and
- *Step 4- select the batch of VMs for migration*.

In each iteration, Step 1 collects the information necessary for the upgrade of the IaaS resources by creating or updating the RG. This graph is created in the initial iteration and then updated in each subsequent one. The inputs for this step in the initial and in the subsequent iterations, while similar, are not the same. In the initial iteration, the RG is created according to the current configuration of the system, the requested change sets, and the infrastructure component descriptions provided by vendors. In the subsequent iterations, the upgrade request model including the state of ongoing upgrade requests and the *upgrade iteration report* indicating the results of the previous iterations are used as additional inputs. Among others the upgrade iteration report indicates the failure of upgrade actions of the previous iteration, as well as the failed and isolated-only resources, based on which undo/retry of the operations can be initiated as necessary.

As mentioned earlier, the configuration of the system may also change between two subsequent iterations independent of the upgrade process due to live migrations, failures, and scaling in/out. Thus, in each iteration the RG is updated based on the current configuration of the system. The RG update also takes into account any new upgrade request and other updates to the upgrade request model.

In Step 2, the resources that need to be upgraded at the same time are identified based on their dependencies. The vertices of these resources are merged to coarsen the RG into an upgrade *Control Graph* (CG), where each vertex represents a resource group grouping of one or more resources that need to be upgraded at the same time. The CG is created in the initial iteration and updated in the subsequent ones to reflect the updates of the RG. A vertex of the CG maintains all the information of the vertices of the RG from which it was formed. For example, the actions-to-execute attribute of the resource group is formed by merging per execution level the actions-to-execute attributes of the resources of the group. In subsequent steps the resources to be upgraded in the current iteration will be selected based on the resource groups and their dependencies of the CG.

Thus, in Step 3, first the IaaS resource groups that can be upgraded without violating any of their dependency requirements are selected to form an initial batch. To satisfy SLA constraints this initial batch may be reduced resulting in a final batch. Accordingly, a runtime upgrade schedule is generated consisting of the upgrade actions for the final batch. This upgrade schedule is sent to the upgrade engine for execution. After the execution of the upgrade schedule, the upgrade engine provides feedback, including any failed upgrade action, to the upgrade coordinator. Based on this feedback, the upgrade coordinator may create a new runtime upgrade schedule still within the same iteration to handle the failed upgrade actions at the resource level, i.e. to bring them into a stable configuration. Once resource level actions are not appropriate or necessary for the given iteration, the upgrade coordinator proceeds to Step 4.

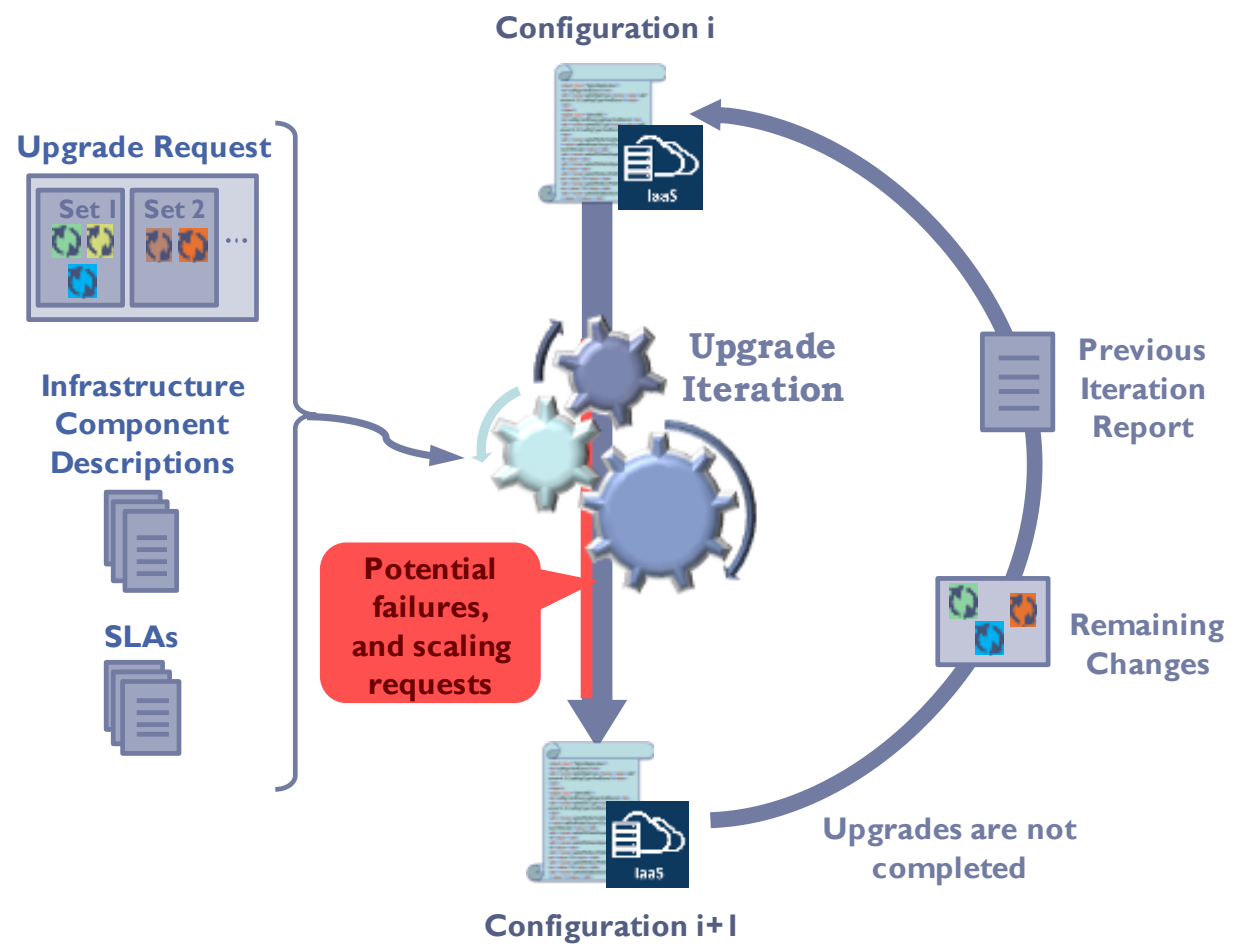

Fig. 4. The iterative process of the IaaS upgrade approach

In Step 4 the VMs hosted by the infrastructure are considered. Since the upgrade of both the VM supporting storage/controller resource and the hypervisor affect the compute hosts on which the VMs are hosted, while they are upgraded the IaaS compute hosts are partitioned into an old and a new partitions. In such situation, a batch of VMs may be selected in this step for migration and possibly upgrade. If these upgrades do not necessitate VM upgrade, in this step a selected batch of VMs is migrated from the old partition to the new one as appropriate. If VM upgrade is also necessary due to incompatibilities between the versions, then the VMs are also upgraded in this process. The selection of the batch of VMs takes into account the results of Step 3. To respect application level redundancy, we can impact at a time only a limited number of VMs (one per anti-affinity group or as appropriate for the SLAs). This means that the selected batch of VMs might need to be upgraded/migrated in sub-iterations within an iteration. In this case, the upgrade coordinator generates an upgrade schedule for each sub-iteration. As in Step 3, the upgrade coordinator sends each schedule to the upgrade engine for execution and based on feedback received generates the next schedule. If an upgrade action fails, the new upgrade schedule also includes the actions reversing the effects of completed upgrade actions for the failed action. The process continues until all the VMs in the selected batch have been handled. If the compute hosts are not partitioned, this step is skipped all together.

At the end of each iteration the upgrade coordinator updates the upgrade request model, the RG and the CG, and generates the upgrade iteration report to reflect the execution results of all schedules within that iteration. The upgrade iteration report indicates the failed and/or isolated-only resources and failed undo units of the iteration. Based on this report, in the subsequent iteration(s) the upgrade coordinator can issue the retry or undo operations as appropriate at the system level considering all the relevant dependencies including those defined by the grouping of requested changes in the upgrade request.

If new upgrade requests are issued during an iteration, they will be considered in subsequent iterations as applicable. That is, the proposed upgrade management framework supports continuous delivery. The upgrade process terminates when all upgrade requests indicated in the upgrade request model have been handled and no new upgrade request has been received. This means that all change sets of all the upgrade requests received have been applied successfully or undone unless their target resources failed.

## IV. Details of the IaaS Upgrade Approach

Hereafter, we elaborate more on each of the steps of our approach.

### *Step 1. Creating/updating the resource graph*

As we mentioned earlier, the upgrade requests received from the administrator are processed and aggregated into the upgrade request model, which is used as input to create and update the RG.

For creating the RG in the first iteration, all existing resources (i.e. vertices) and dependencies (i.e. edges) are extracted from the current configuration of the system. Their parameters are derived from the system configuration (e.g. resource-id) and the upgrade request model (e.g. modification-type). The resources to be added are determined from the change sets in the upgrade request model. For them the parameters and dependencies are derived from the upgrade request model and the infrastructure component descriptions provided by the vendor. To satisfy the requirements indicated by the vendors, each change set is verified for completeness and any missing/detailed changes are added to the upgrade request model. These are also reflected in the RG. In this process each change set is assigned to a unique undo unit.

The actions-to-execute attribute of each resource is determined using the infrastructure component descriptions kept in the resource upgrade catalog. If the required upgrade actions cannot be applied to a resource in a single iteration due to ordering constraints with respect to other resources, the upgrade actions are split into different execution-levels to enforce the ordering.

To avoid the interaction between resources of incompatible versions during their upgrade, the upgrade of dependent resources with incompatibilities need to be carried out using an upgrade method, which handles appropriately these incompatibilities. For this, the upgrade coordinator first identifies such resources in the RG and then groups them into an upgrade unit. For each upgrade unit an appropriate upgrade method is associated automatically considering the potential incompatibilities along the dependencies of its resources. The two basic upgrade methods used are: split mode and rolling upgrade. Split mode is typically used in case of incompatibilities and rolling upgrade otherwise. Other upgrade methods may be used as well depending on the situations.

To update the RG in a subsequent iteration, first the RG is updated to reflect the current configuration of the system with respect to any changes that occurred in the system.

The upgrade iteration report of the just completed iteration helps in identifying any retry and system level undo operations needed. The RG is updated as follows: Resources with the number of failed upgrade attempts equal to the retry thresholds are isolated. Whenever, the number of isolated-only and failed resources for an undo unit reaches the undo threshold, all the changes already applied to the resources of the undo unit must be undone. In addition, the RG is updated to include upgrade actions for an undo operation for any undo unit whose upgrade did not completed within the time limit indicated as max-completion-time. This is measured from the time of the time stamp of the upgrade request with the corresponding change set. These undo units and the associated change sets are also marked as failed.

While updating the RG with respect to an undo operation, the actions-to-execute attributes of all the affected resources (excluding the failed resources) in the failed undo unit are adjusted so that they will be taken to the undo-version indicated for the resources. These undo actions are organized into the first execution level of the resources so that they will be executed first. Since there might be upgrade actions associated with other change sets in the actions-to-execute attributes of these resources which were not completed yet, they need to be adjusted as well. For this, the upgrade actions of other execution levels of the resources are re-evaluated with respect to the potentially new source and target versions as well as the upgrade actions are updated based on the component descriptions in the catalog. Isolated-only resources which are at the undo-version are released from isolation, otherwise an attempt is made to take them to the undo-version. If this attempt fails, they are marked as failed resources.

As mentioned earlier, new upgrade requests are added to the upgrade request model and then to the RG. New upgrade requests may be targeting resources that are part of pending change requests. Such new upgrade request may also result in new incompatibilities. To identify these, a separate graph similar to the RG is used: The *New Request Graph (NRG)*. This graph is created only from the new upgrade requests without considering any ongoing upgrades. The upgrade coordinator extracts from the component descriptions the upgrade actions for the new change sets and organizes them into execution levels as required. Next, any newly introduced incompatibilities are identified and corresponding new upgrade units are created in the NRG. The upgrade coordinator uses this NRG to update the RG as follows: With respect to the actions-to-execute attributes of resources already in the RG, a new execution level for each execution level is created and appended in the NRG. The newly added execution levels are associated with the upgrade units identified in the NRG.

### *Step 2. Grouping the IaaS resources for upgrade*

Some dependencies between resources require that these resources are upgraded at the same time in a single iteration. To facilitate the coordination of the upgrade of these resources, the upgrade coordinator coarsens the RG, into the CG. In the CG each vertex represents a resource group, i.e. an individual resource or a group of resources of the RG to be upgraded at the same time. Here we provide more details on the creation/update of the CG:

*Dependency based edge contraction:* During the upgrade of a container its contained resource(s) experience an outage in addition to the outage during their own upgrade. Likewise, during the upgrade of constituent resources, their composite resource experiences an outage. To reduce the outage time, resources with container/contained and resources with composition dependencies should be upgraded at the same time in a single iteration. Thus, the upgrade coordinator contracts the edges representing such dependencies in the RG to merge the vertices representing these resources into a single vertex of the CG. A vertex in the CG, representing a resource group of the RG, will have the same dependencies to (groups of) other resources as the resources of the merged vertices of the RG except for the container/contained and the composition dependencies, which are eliminated by the edge contraction.

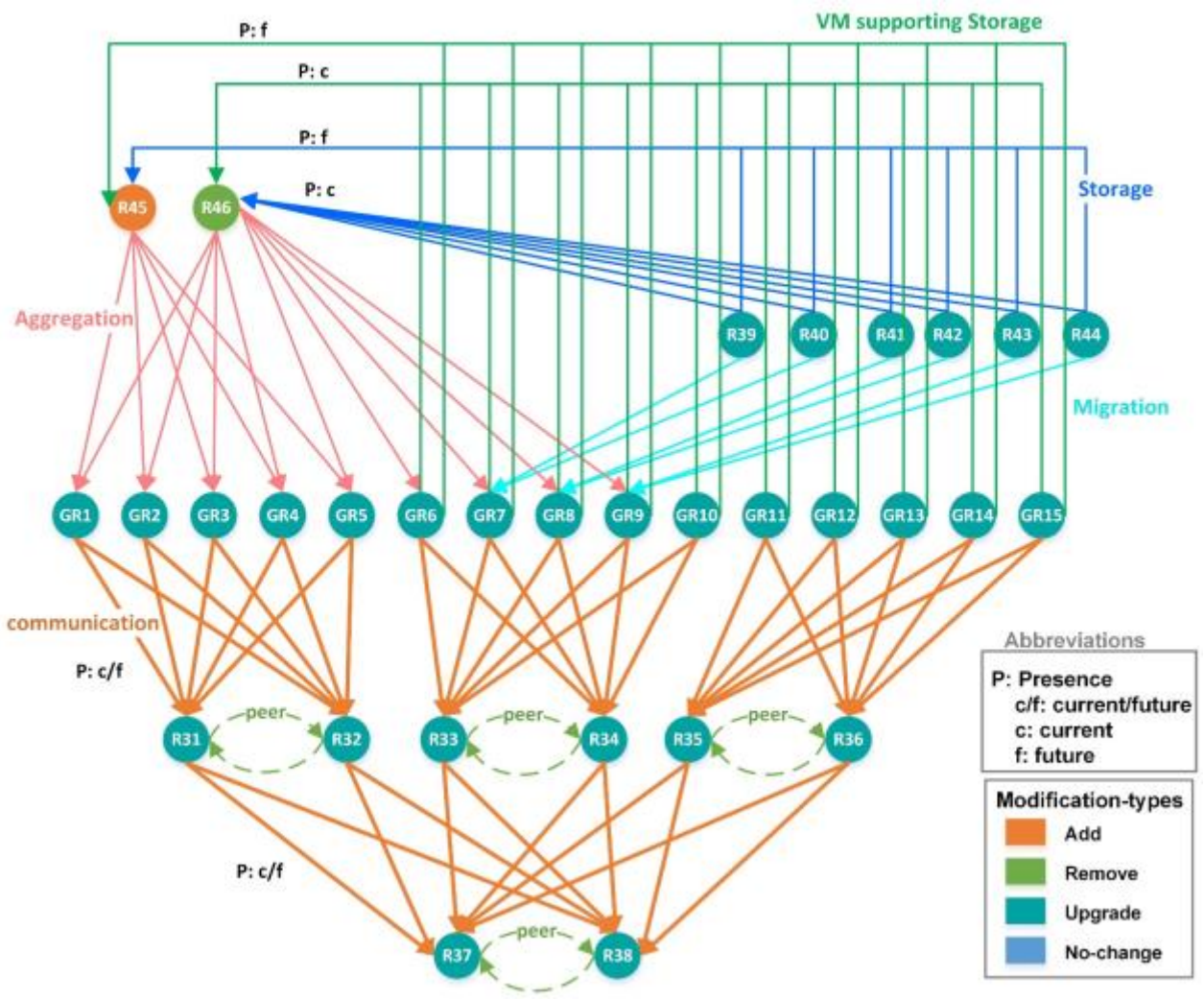

Fig. 5. Upgrade control graph for the illustrative example

Fig. 5 shows the CG corresponding to the RG of Fig. 2 for our illustrative example in Fig. 1. An edge contraction of this type was applied to the vertices of the RG representing the resources R1, R16, R47, R48, R49, and R50 to coarsen them into vertex GR1 of the CG. GR1 keeps the edge R16 had toward R45, R46, R31 and R32 in the RG. Note that in Fig. 5, the upgrade related parameters of the CG are not shown.

*Upgrade method based vertex contraction:* Some upgrade methods avoid incompatibilities by upgrading resources at the same time in a single iteration. The upgrade coordinator performs vertex contraction for such resources based on the upgrade methods associated with the first execution-level in their actions-to-execute attribute. In case of a vertex contraction, the resulting vertex of the CG will have the union of all dependencies that the resources of the group had in the RG. For example, the vertices representing the resources of an upgrade unit to be upgraded using the split mode upgrade method, will be contracted according to the sub-partitioning of the upgrade unit for the split mode. This allows the proper coordination of the upgrade of the resources without introducing incompatibilities.

The CG is also updated continuously to maintain consistency with the RG.

### *Step 3. Selecting the batch of IaaS resources for upgrade*

In this step, the batch of IaaS resources to be upgraded in the current iteration is selected considering both the existing dependencies and the SLA constraints. Since VMs represent the service the IaaS cloud system provides, they are handled separately in Step 4 by considering different criteria.

In Step 3 first, the VMs are consolidated on the compute hosts as much as possible to free up some hosts. In particular, if VM supporting storage and controller resources need to be upgraded in an incompatible way, the VMs are evacuated from the physical hosts in common between the sets of $M_{Storage}$ and $M_{Compute}$, to accommodate the PPU method as much as possible. During VM consolidation, the availability constraint, inferred from the anti-affinity grouping, are respected by migrating only one VM at a time from each anti-affinity group. After consolidation, the RG and the CG are updated accordingly.

To handle the dependencies during the upgrade, first the initial batch ($G_{batch}$), i.e. the resource groups that can be upgraded in the current iteration without violating any of their dependencies are identified using the CG. To do so in a systematic way, the upgrade coordinator initializes $G_{batch}$ as the union of the set of CG vertices with remaining changes (i.e. modification-type of "Upgrade", "Add", "Remove") and the set of CG vertices with deactivated status (i.e. with activation-status out of service).

Next, the vertices, which cannot be upgraded in the current iteration due to some dependencies are eliminated from $G_{batch}$. To do so we have defined a set of rules, referred to as elimination rules. The upgrade coordinator uses the elimination rules to identify the non-suitable candidates in $G_{batch}$ and remove them.

The elimination rules are based on:

- the modification-type of the resources,
- the upgrade method associated with the upgrade unit of the first execution level in the actions-to-execute attribute of the resources,
- the incompatibility-factor and presence attributes of the dependencies of the resources,
- the activation-status of the resources, and
- the availability of additional resources required as prerequisite for the related upgrades.

These elimination rules guarantee:

- the enforcement of compatibility requirements of sponsorship dependencies between resources,
- the availability of services provided by peer resources,
- the satisfaction of the resource requirements of the PPU method,
- keeping the VM service available,
- the satisfaction of dependency requirements (i.e. before removing a resource from the system, and before adding a resource to the system), and
- the correct ordering of resource upgrades with respect to the upgrade method associated with the upgrade unit of the first execution-level in the actions-to-execute attribute of the resources.

Here we discuss only one of the elimination rules in details, which guarantees the satisfaction of the resource requirements of the PPU method when used for upgrading a VM supporting storage/controller resources. As mentioned earlier, to maintain in parallel both the old and the new configurations of the VM supporting storage/controller resources, additional resources may be required. If these cannot be provided using available resources, the administrator is asked to provide additional resources. Until these resource requirements are not satisfied, all the resources with changes related to the upgrade of the VM supporting storage/ controller resources are eliminated from $G_{batch}$.

In our illustrative example given in Fig. 1, the PPU method is used to upgrade the VM supporting virtual shared storage from VSAN to Ceph as the new (Ceph) and the old (VSAN) versions of the virtual shared storage are incompatible. To keep the continuity of the VM supporting service (e.g. VM live migration and failover) during the upgrade, the old configuration of the virtual shared storage (i.e. VSAN) has to remain operational until the new configuration (i.e. Ceph) is ready for use. In addition, the compute hosts hosting the VMs need to be partitioned into those compute hosts compatible with the old version of the virtual shared storage (old partition) and those compute hosts compatible with the new version of the shared storage (new partition). To complete this upgrade, data conversion is also necessary, and it is performed as the VMs are migrated from the old partition to the new. Once all the VMs have been migrated as well as the data migration have been completed, the old configuration of the virtual shared storage can be safely removed.

To guarantee the continuity of services supporting VM operations during the upgrade of the shared storage, the minimum resource requirements for both the old and the new virtual shared storages with respect to their configurations and the data stored, need to be satisfied. For this reason, there have to be enough physical storage hosts to keep the old configuration of the storage alive while bringing up the configuration of the new. This elimination rule evaluates whether the current system has enough storage hosts, using (1).

$$|M_{storage} - M_{usedCompute}| \geq \max(MinHostReqConf_{oldStorage}, MinHostReqCap_{oldStorage}) + \max(MinHostReqConf_{newStorage}, MinHostReqCap_{newStorage}) \quad (1)$$

Please refer to Table A for the notation used.

$|M_{Storage}$-$M_{usedCompute}|$ represents the number of storage hosts that are not in use as compute hosts. This number should be equal to or greater than the minimum number of hosts required to support both the old and the new storage configurations during the upgrade. If (1) is satisfied, the resources with upgrade actions related to the undo unit associated with virtual storage upgrade remain in $G_{batch}$. Otherwise, applying the elimination rule will remove these resources from $G_{batch}$ as non-suitable candidates. Since the same check is performed in each subsequent iteration, whenever the required number of storage hosts becomes available to fulfill this requirement, these resources will remain in the $G_{batch}$ as suitable candidates. Note that as the upgrade proceeds the number of available resources may change due to failures or scaling operations on compute hosts, but also if additional hosts are provided. Thus, in any iteration when (1) is not satisfied, this elimination rule will remove from $G_{batch}$ the resources related to the upgrade of VM supporting storage/controller resources (i.e. their upgrade will be paused) until the required resources will become available (again).

After applying all elimination rules, the vertices remaining in the $G_{batch}$ represent the initial batch. However, this selection does not consider yet the dynamicity of the IaaS cloud; i.e. SLA violations may still occur if all these resource groups are upgraded in the current iteration. Namely, considering potential failovers and scale-out requests during the iteration only a smaller number of compute hosts can be taken out of service. Thus, with these considerations we select a final batch of resource groups from the initial batch.

We estimate the potential scale-out requests in each iteration based on the time required to upgrade and to recover from possible failures (i.e. using resource level undo) for the initial batch, in which the resources are upgraded in parallel. In each iteration, the $G_{batch}$ may contain different resources, hence in each iteration we calculate this required time ($T_i$) as the maximum time required among resources in $G_{batch}$. For each resource the time required to upgrade and to recover from a failure is determined based on the time estimate for installation/removal provided by each vendor in the related infrastructure component description.

Using this $T_i$ the maximum scaling adjustment requests per tenant ($S_i$) during the upgrade of $G_{batch}$ in iteration *i* is calculated according to (2).

$$S_i = max(\, s_n * \left\lceil \frac{T_i}{c_n} \right\rceil) \tag{2}$$

Where $s_n$ is the scaling adjustment per cooldown period $c_n$ of the $n_{th}$ tenant. Since tenants may have different scaling adjustment and cooldown time values, we take the maximum scaling adjustment among them as $S_i$ and by that we handle the worst case scenario. This calculation is also valid for a single iteration only and it is recalculated for each iteration since in each iteration different resources may remain in the $G_{batch}$, as well as tenants may be added and/or removed.

We calculate the maximum number of compute hosts that can be taken out of service ($Z_i$) in each iteration using (3).

$$Z_i = |M_{computeForOldVM} - M_{usedComputeForOldVM}| - Scaling\ Re\,s\,v_{forOldVM} - Failover\ Re\,s\,ev_{forOldVM} \tag{3}$$

Where $|M_{ComputeForOldVM}$ - $M_{usedComputeForOldVM}|$ is the number of compute hosts that are not in use and are eligible to provide compute services for tenants with VMs of the old version (i.e. compatible with the old configuration of VM supporting storage/controller resources or old hypervisor). $FailoverResev_{forOldVM}$ is the number of compute hosts reserved for failover for VMs of the old version. This number is equal to the number of host failures to be tolerated during an iteration (F), when there are VMs of the old version on hosts belonging to $M_{ComputeForOldVM}$ (i.e. $M_{usedComputeForOldVM}$ is not zero); otherwise F will be zero. F can be calculated based on the hosts' failure rate and a probability function as in [22] which estimates the required failover reservations for period $T_i$. $ScalingResv_{forOldVM}$ is the number of compute hosts for scaling reservation of tenants with VMs of the old version and it is calculated using (4).

$$Scaling\ Re\,s\,v_{forOldVM} = S_i * \left\lceil \frac{A_i}{K} \right\rceil \tag{4}$$

Where $A_i$ indicates the number of tenants with VMs of the old version only and who have not reached their $max_n$, the maximum number of VMs, therefore may scale out on hosts compatible with the old version of the VMs.

Whenever $M_{usedComputeForOldVM}$, the set of compute hosts in use with the old version is empty, the maximum number of compute hosts that can be taken out of service in the iteration becomes equal to the set of hosts belonging to $M_{computeForOldVM}$.

Note that if the compute hosts of IaaS cloud system are not partitioned into old and new partitions, the above calculations are applied to all compute hosts (as opposed to those hosting old VMs) and all VMs as there is no need to consider the compatibility of VMs and compute hosts. Without incompatible partitions there is also no need for Step 4 in the given iteration.

Resource groups can be selected from the initial batch such that their total number of affected compute hosts for the duration of $T_i$ is not more than $Z_i$. To select this final batch $G_{Fbatch}$ from the initial batch $G_{batch}$, resource groups which can return to service immediately after their upgrade are distinguished (based on their associated upgrade methods) from those that need to be kept deactivated due to potential incompatibilities. The latter ones impact subsequent iterations by contributing to $Z_{i+1}$ at least, hence they require additional resources to prevent SLA violations. We can only upgrade a limited amount of such resource groups in an iteration depending on the availability of resources that can be taken out of service for a period longer than $T_i$. The availability of such resources depends on, for example, the cloud provider's booking strategy. They are not available when the cloud provider commits to provide more VMs than the actual capacity of the system, referred to as overbooking [23][24]. Hence, we expect some policy from the cloud provider guiding such a selection. The cloud provider may indicate resources dedicated to upgrades, which are not counted in the system capacity therefore using them for the upgrade of resource groups, which remain deactivated without impact on subsequent iterations. The amount of such dedicated additional resources is used as upper bound for the selection of such resource groups.

The upgrade coordinator selects such a final batch $G_{Fbatch}$ and generates the corresponding upgrade schedule. This upgrade schedule includes the upgrade actions of the first execution-level of the actions-to-execute attribute of each resource group in $G_{Fbatch}$. The generated schedule is sent to the upgrade engine for execution. After execution, the upgrade engine sends back to the upgrade coordinator the results.

Note that some upgrade actions may require the application of some prerequisite actions beforehand and some wrap-up actions afterwards. For example, the prerequisite action for upgrading a physical host could be the evacuation of VMs from the host. Accordingly, the wrap-up actions would bring the VMs back to the upgraded host. These prerequisite/wrap-up actions are associated with the deactivation/activation of the resources to be upgraded. Therefore the upgrade coordinator checks the upgrade method associated with the upgrade unit of the resources in the final batch. If the upgrade method requires the activation or deactivation of a resource and therefore it requires such prerequisite and wrap-up actions, the upgrade coordinator includes them in the upgrade schedule: the prerequisite actions before the upgrade actions of that resource and wrap up actions after them.

If the upgrade actions of a resource in the final batch were executed successfully, the first execution-level is removed from its actions-to-execute attribute together with its executed upgrade actions. Thus, the next execution-level becomes the new first execution-level of the actions-to-execute attribute. The modification-type of the resource is adjusted according to the upgrade actions of this new first execution-level.

For a resource with a failed upgrade action, the counter of failed attempts is incremented, but the actions-to-execute attribute remains unchanged. To recover from the failure and to bring the resource back to a stable configuration, a new upgrade schedule is created from the undo actions of the completed upgrade actions within the failed attempt to revert their effect, that is, a resource level undo operation is constructed. This upgrade schedule is given to the upgrade engine right away for execution. If this operation fails as well, the resource is isolated and marked as a failed.

At the end of the step, the upgrade request model, the RG and the CG are updated according to the results.

*Step 4. Selecting the batch of VMs for migration*

This step is only necessary when the compute hosts are separated into two incompatible partitions due to the upgrade of the VM supporting storage/controller and/or the hypervisors hosting VMs, and therefore the VMs need to be migrated between them while they may also need to be upgraded. For example, when the PPU method is used to handle the incompatibilities of the VM supporting storage/controller resources.

Before VMs of the old version can be upgraded and migrated to the hosts compatible with the new VM version, the new configuration of the VM supporting storage/controller resource needs to be completed. If the new configuration is not ready the VM migration/upgrade is delayed to a subsequent iteration, when it is re-evaluated. In case of incompatibilities due to hypervisor upgrade, this step can be started after a successful upgrade of at least one hypervisor.

We calculate the number of VMs ($V_i$) that can be migrated and if necessary, upgraded in the current iteration $i$ using equation (5).

$$V_i = (\left|M_{computeForNewVM} - M_{usedComputeForNewVM}\right| - Scaling\,Resv_{forNewVM} - Failover\,Resev_{forNewVM}) * K' \quad (5)$$

Where $M_{computeForNewVM}$ is the set of hosts that are eligible to provide compute services for tenants with VMs of the new version, $M_{usedComputeForNewVM}$ is the set of in-use hosts that are eligible to provide compute services for tenants with VMs of the new version, $FailoverResev_{forNewVM}$ is the number of hosts reserved for any failover for upgraded (new) VMs. $FailoverResev_{forNewVM}$ is calculated similarly to the failover reservation for tenants with VMs of the old version, i.e. F as mentioned in Step 3, but for the period of time required for upgrading $V_i$ number of VMs. $ScalingResv_{forNewVM}$ is the number of hosts reserved for scaling for the tenants with upgraded (new) VMs, and $K'$ is the new host capacity in terms of VMs after the upgrade. Here, $ScalingResv_{forNewVM}$ is calculated similar to (4) for the tenants with VMs of the new version who have not reached their $max_n$ (their maximum number of VMs). They may only scale out on hosts compatible with VMs of the new version. Note that a new scaling adjustment per tenant have to be calculated similar to (2), while considering the time required to migrate, upgrade and recover from possible failures for $V_i$ number of VMs potentially through multiple sub-iterations as discussed below.

Considering the application level redundancy, in the worst case we can migrate (and upgrade) only one VM per anti-affinity group at a time. Therefore, we may need to upgrade the $V_i$ VMs in several sub-iterations. Thus, the time required to migrate, upgrade, and recover from possible failure for $V_i$ number of VMs depends on the number of sub-iterations and the time required for a single VM. In each sub-iteration $j$, one VM is selected from each anti-affinity group with VMs of the old version. The batch of sub-iteration $j$ will be $W_{ij}$. In order to speed up the upgrade process, we use two criteria for selecting the anti-affinity groups and their VMs for the upgrade:

1) To free up more hosts, anti-affinity groups of tenants with the highest number of VMs of the old version are preferred, and
2) To minimize the number of VM migrations, VMs of hosts with more VMs from the preferred anti-affinity groups are selected for the migration.

VMs of selected anti-affinity groups can belong to the tenants that did not have upgraded (new) version VMs yet. The number of such tenants was considered in the scaling reservation calculation for the old partition and not for the new. However after migrating (and upgrading) their VMs, they will be required to scale-out on hosts compatible with VMs of the new version. Thus, before performing the migration (and upgrade) of the selected VMs, $ScalingResv_{forNewVM}$ must be re-evaluated to determine if it is sufficient for scaling-out of such tenants as well. The batch of VMs for each sub-iteration may have to be re-adjusted accordingly. This re-adjustment is based on the number of not-in use hosts (compatible with VMs of the new version), the newly calculated scaling reservation, and failover reservation.

After the upgrade coordinator selects the VMs for the migration/upgrade, it creates a schedule per sub-iteration and provides the schedule to the upgrade engine for execution. After the execution of each sub-iteration, the upgrade engine returns the results to the upgrade coordinator. As for resources, the actions-to-execute attribute of VMs successfully migrated/upgraded is updated by removing the first execution level. For VMs with failed attempts, a new schedule is generated to bring a new (version of the) VM in its initial state up on the hosts compatible with the (new) VM version. If the number of migrated/upgraded VMs is less than $V_i$ and there are more VMs for migration/upgrade, the upgrade proceeds to the next sub-iteration. Otherwise, the upgrade proceeds to the next iteration starting with Step 1.

Whenever in Step 3 the final batch of resources ($G_{Fbatch}$) and in Step 4 the batch of VMs ($V_i$) are both empty for an iteration,

the upgrade process is suspended until enough resources become available to continue, e.g. freed up through scaling in. Thus the upgrade coordinator monitors the system configurations for changes and re-evaluates the situation.

## V. IMPLEMENTATION AND VALIDATION

### A. Implementation, Settings and Evaluations

To demonstrate the feasibility of the proposed upgrade management framework in a real deployment, we have implemented a proof of concept (PoC) for the upgrade of compute hosts of an OpenStack [25] deployment. In this PoC, the proposed upgrade management framework is deployed as an upgrade service on the OpenStack controller node to manage the upgrade of compute hosts. As a case study, we considered the upgrade of the hypervisors (QEMU [26]) on the compute hosts in a virtual OpenStack cluster.

We performed some evaluations to demonstrate how our approach handles the SLA constraints of availability and elasticity compared to the traditional rolling upgrade method with different fixed batch sizes. In our evaluation, we considered ten compute hosts, hosting VMs from four tenants with scaling parameters as shown in Table I. For simplicity, we assume that all tenants have the same scaling adjustment and cooldown period configured. Also all VMs of a tenant form a single anti-affinity placement group from which if more than one VMs are taken out of service the tenant experiences an application level outage.

We evaluated two different scenarios (a) when the tenants have their initial number of VMs on the hosts as shown in Fig. 6.a, and (b) as shown in Fig. 6.b. To have a fair comparison of our approach with the rolling upgrade method, in our evaluation scenarios we assumed no incompatibilities during the upgrade of hypervisors, and VMs could be migrated between the old and the new versions of the hypervisors with no need for VM upgrade.

The required measurements for our evaluation have been obtained from a real deployment with six compute nodes. The upgrade scenario for the six nodes was executed ten times, and we considered the average measurements. Accordingly, upgrading a QEMU hypervisor (i.e. executing the compiled binaries of the new version on a compute node) takes on average 41 seconds. Live migrating a VM from the old to the new version of the hypervisor takes on average 23 seconds. The outage introduced during the live migration of a VM (of tiny flavor) by OpenStack takes less than 0.6 seconds according to [27]. Performing the necessary calculations for each iteration of the upgrade takes on average 0.23 seconds for our approach. These measurements have been used in our calculations for the evaluation scenario with ten compute nodes.

| Tenant ID | Tenant | Initial Number of VMs | Min-size | Max-size | New Version |
|---|---|---|---|---|---|
| 1 | | 2 | 2 | 6 | |
| 2 | | 3 | 3 | 7 | |
| 3 | | 3 | 2 | 5 | |
| 4 | | 1 | 1 | 4 | |

TABLE I. SCALING PARAMETERS FOR THE TENANTS IN OUR EVALUATION SCENARIO

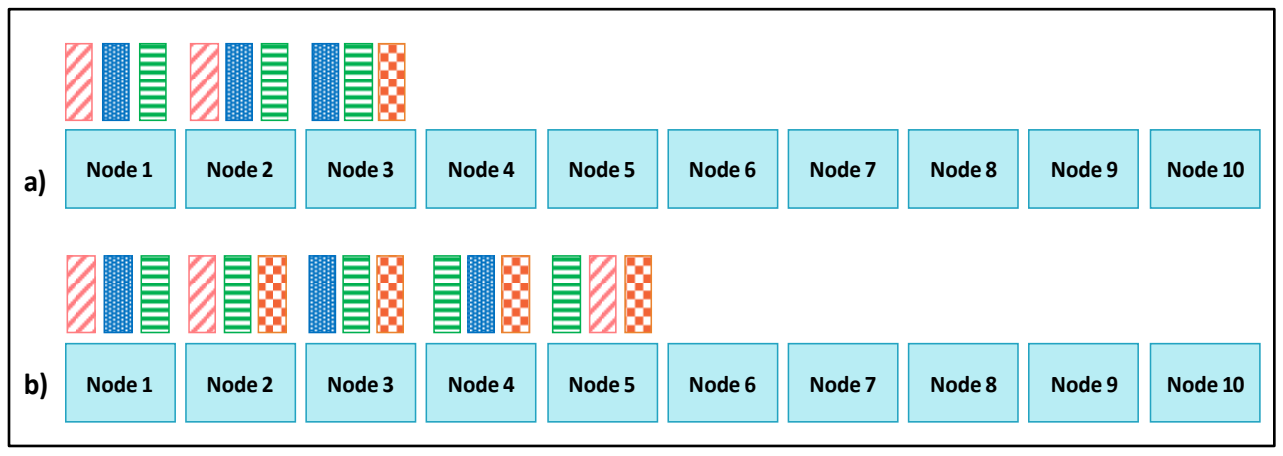

Fig. 6. Two different cases for our evaluation scenario

To evaluate the availability at the application and at the VM levels, we calculated and compared: the total duration of the upgrade of ten compute nodes, the average outage time at the application level for each tenant, and the average outage time of each VM during the upgrade. With respect to SLA violations, we calculated and compared: the number of SLA violations per tenant, the duration of the SLA violation for each breach, the total duration of SLA violation per tenant, and the applicable penalties.

The penalties are formulated in different ways by different cloud providers. In our evaluation, we modified the *proportional penalty* as described in [28], to motivate higher availability. The proportional penalty is a form of delay-dependent penalty, where the penalty is proportional to the occurred delay in providing the required capacity and the difference between a user's provisioned capacity and the expected allocation. It is calculated by multiplying an agreed penalty rate of $q$ (per unit capacity per unit time), the duration of SLA violations, and the difference in the expected and provisioned capacity. To motivate higher availability and to penalize more lower than expected capacity, we make the capacity term of the proportional penalty function quadratic and refer it to as *quadratic proportional penalty*.

We assume that the rolling upgrade is performed with a fixed batch size and that the VMs of the nodes selected for the upgrade are migrated to other nodes prior to the upgrade. The selection of the nodes in each batch of upgrade, the distribution of the VMs on these selected nodes, and the order of their upgrade can result in different outages and SLA violations. Therefore, we performed our assessments considering different batch selections and considered the average result.

### B. Results and Analysis

Fig. 7 and Fig. 8 show the comparison of the total duration of the upgrade using rolling upgrade with fixed batch sizes and our approach for evaluation scenarios (a) and (b), respectively. The results show that in evaluation scenario (a), the duration of the upgrade using our approach is shorter than using the rolling upgrade method with fixed batch size of one, two, and three, while it is comparable with the duration of upgrade using batch size of four. For the evaluation scenario (b), the duration of upgrade using our approach was shorter than the rolling upgrade method with fixed batch size of one and it was comparable to the duration of rolling upgrade with batch size of two. In this scenario, depending on the selection of in-use or not in-use nodes for the batch and the upgrade order of the batches, the VMs may be migrated once, twice, or three times during a rolling upgrade. This impacts the total duration of the upgrade, as well as the outage time at the application and VM levels.

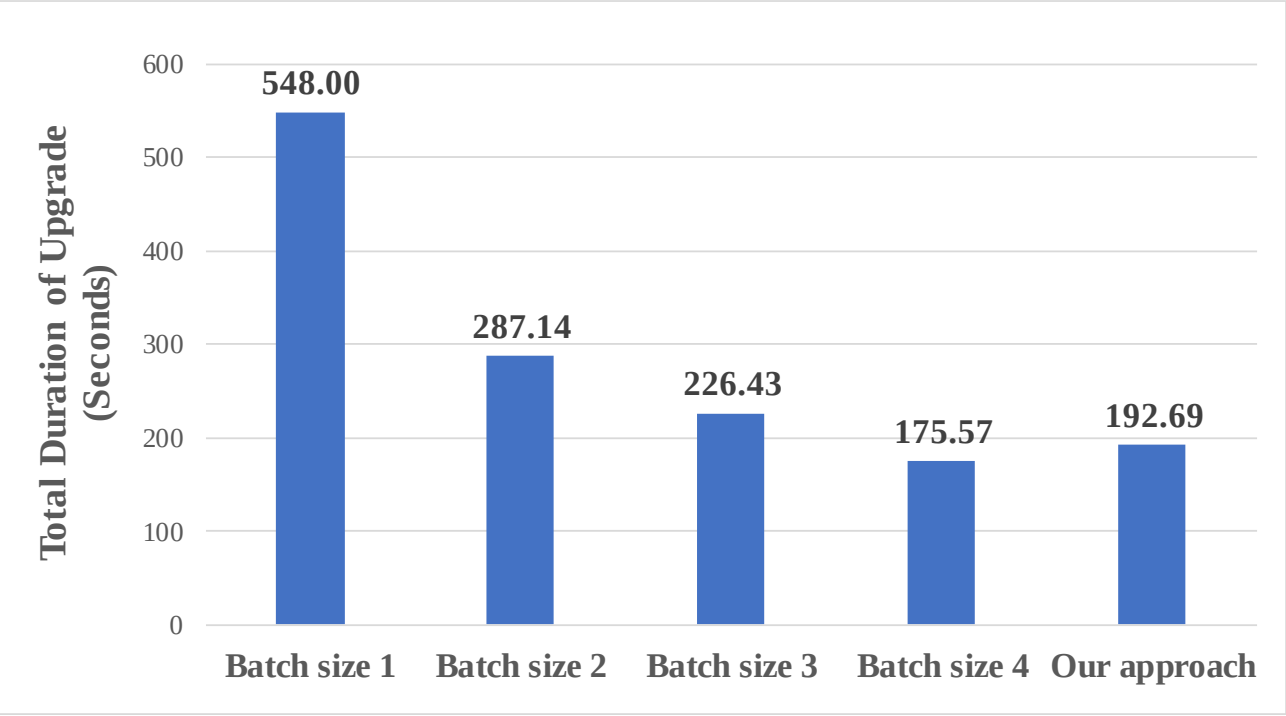

Fig. 7. Comparison of the total duration of the upgrade for evaluation scenario (a)

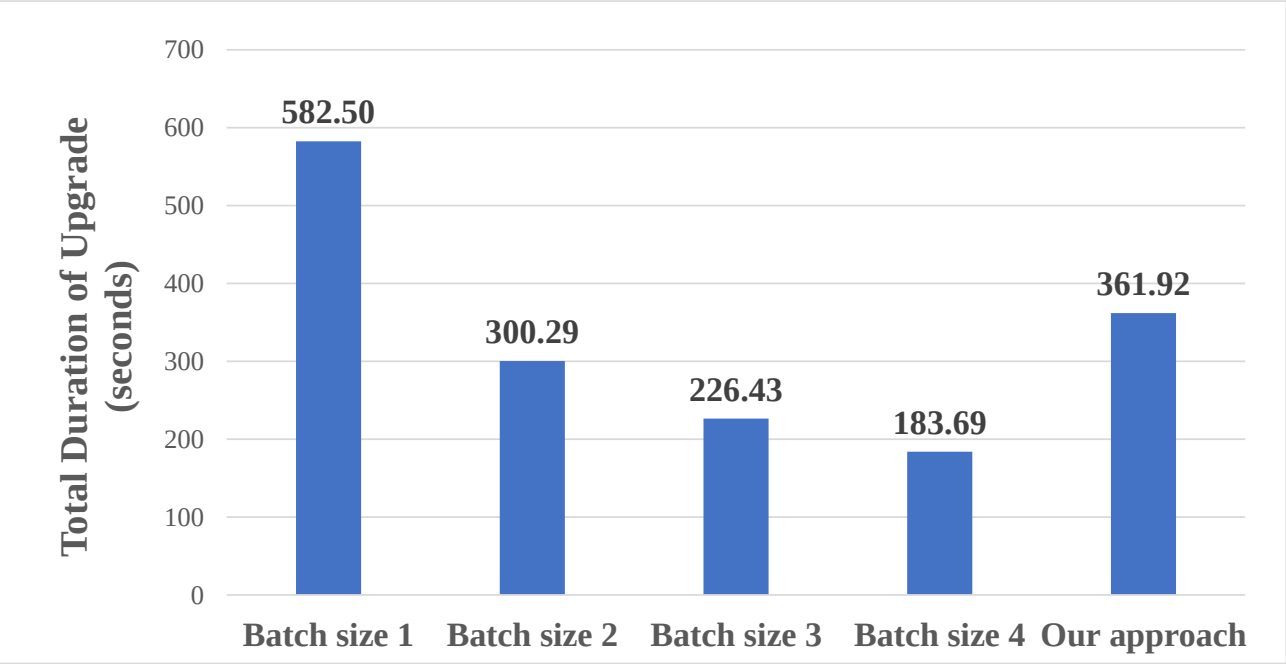

Fig. 8. Comparison of the total duration of the upgrade for evaluation scenario (b)

The duration of the outage introduced at the application level depends on the duration of the outage of each VM during live migration (i.e. 0.6 seconds) and the number of VMs migrated. This is valid for each batch as long as there are enough nodes available to host the evacuated VMs.

Fig. 9 shows the comparison of the average outage at the application level for evaluation scenario (a) and Fig. 10 shows the comparison for evaluation scenario (b) considering different cases: the average outage time for impacted tenants is for the tenants that are impacted at the application level during the upgrade. While, the average outage time per tenant is the average outage time considering all tenants, whether they are impacted or not. In case of rolling upgrade with batch size of one, similar to our approach, tenants do not experience any outage at the application level, except for tenant 4 in evaluation scenario (a). This is because in scenario (a), tenant 4 has only one VM, meaning that it is not configured HA at the application level.

Since in rolling upgrade we assume that the nodes are selected according to the batch size regardless of their usage state, a VM may be migrated between one to three times. Accordingly, the application level outage experienced by tenant 4 in evaluation scenario (a) may be 0.6, 1.2, or 1.8 seconds using rolling upgrade method with batch size of one. Since in our approach we upgrade not in-use nodes before in-use nodes, tenant 4 will only experience an application level outage of 0.6 seconds as shown in Fig. 9. Note that if a similar rule is followed while applying the rolling upgrade method with batch size one, the application level outage for evaluation scenario (a) will also be 0.6 seconds, similar to our approach. If tenant 4 had more than one VM as in scenario (b) of Fig. 10, it would not experience any outage for either cases.

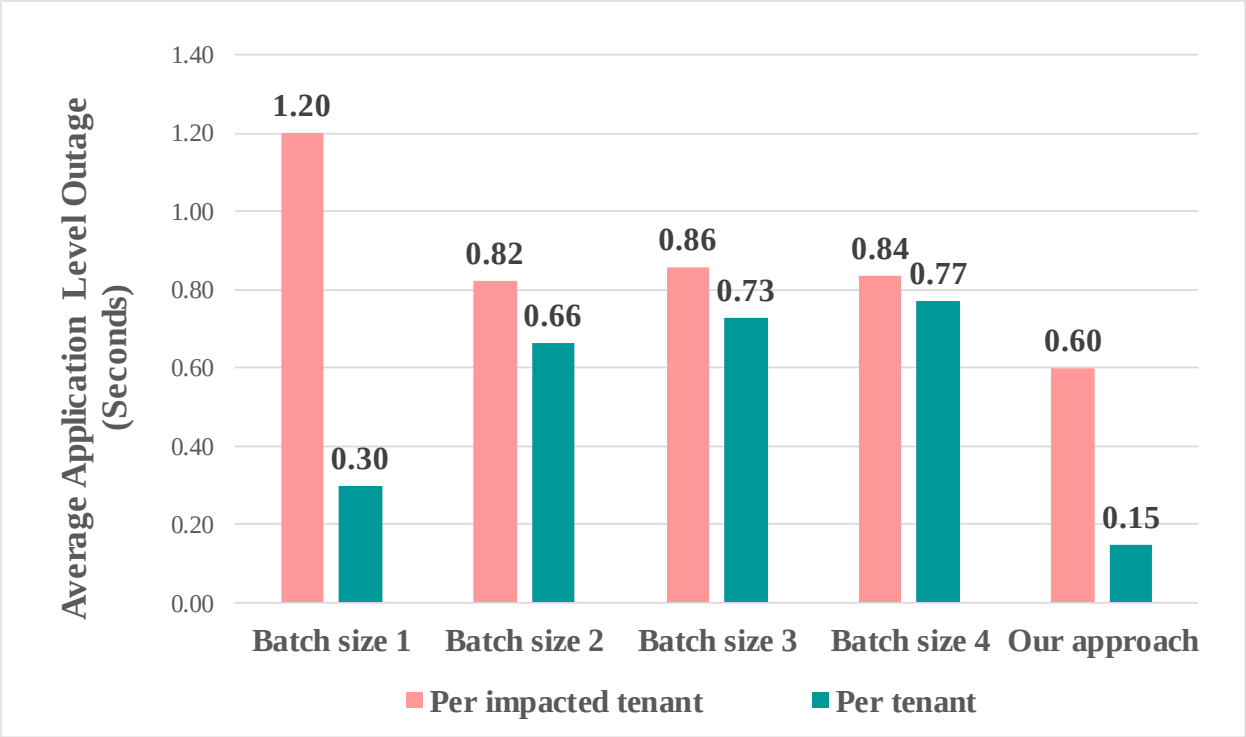

Fig. 9. Comparison of the average outage at the application level for evaluation scenario (a)

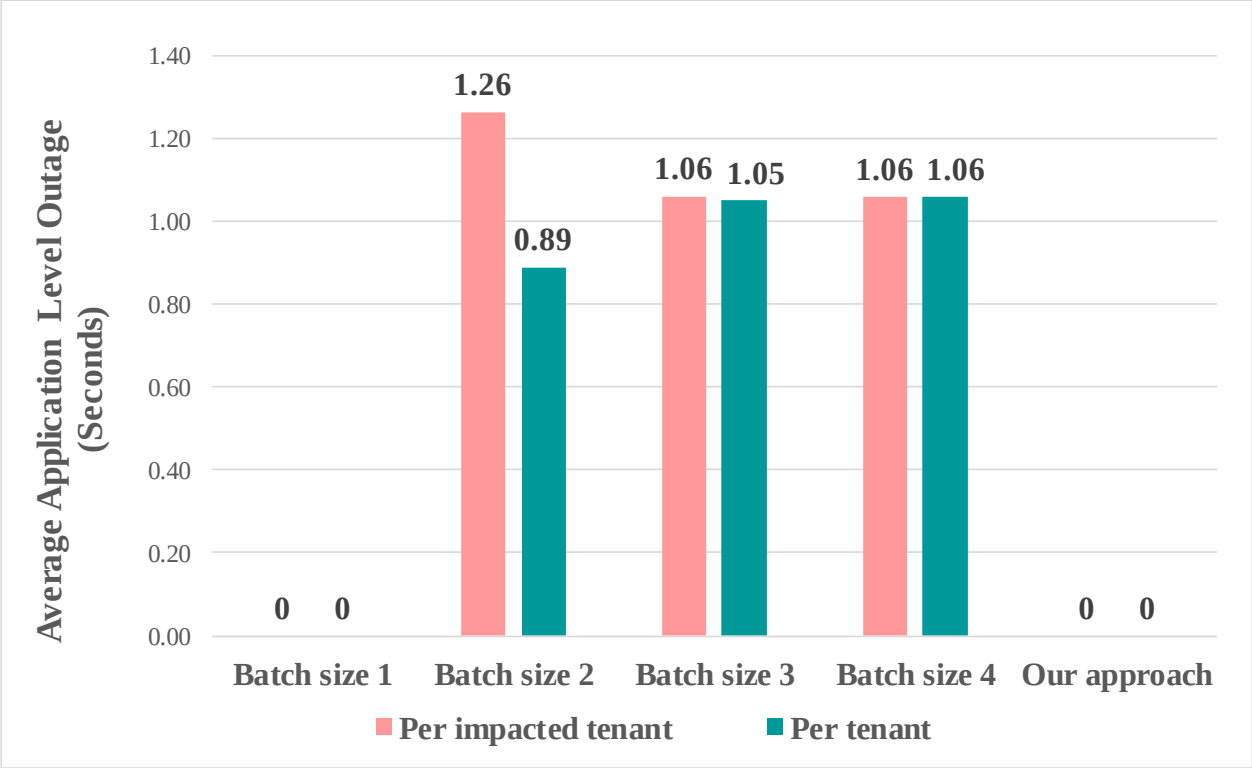

Fig. 10. Comparison of the average outage at the application level for evaluation scenario (b)

As the batch size of the rolling upgrade method increases, the average outage time at the application level increases as well. This is due to the increase of the probability of selecting for the batch, nodes that host VMs from the same anti-affinity group. Note that although in our approach the batch size can be more than one node, we migrate only one VM at a time from an anti-affinity group, which prevents outages at the application level. Therefore, for the tenants that are configured HA at the application level, our approach does not introduce any outage at the application level.

Considering VM level outages, whether using a rolling upgrade method or our approach, for both scenarios of (a) and (b), each VM experiences an outage during its migration. Fig. 11 presents the average outage time of each VM for scenarios (a) and (b). As the results indicate, a VM experiences less outage using our approach, compared to the rolling upgrades, where based on the upgrade order of in-use or not in-use nodes, a VM may be migrated more than once, which impacts the total outage time per VM.

During upgrade whenever VMs experience outage we consider that SLA violations occur. This is because the current number of VMs for the tenants drops below the required number of VMs. Table II and Table III demonstrate the comparison of SLA violations during upgrade scenarios (a) and (b), respectively. In our evaluation we calculated the number of times SLA violations occur for each tenant and we considered the average as the number of SLA violations per tenant. The number of impacted VMs per tenant in each SLA violation indicates how many VMs are impacted. For the average total duration of SLA violations per tenant, we calculated the total time of SLA violations for each tenant and considered their average.

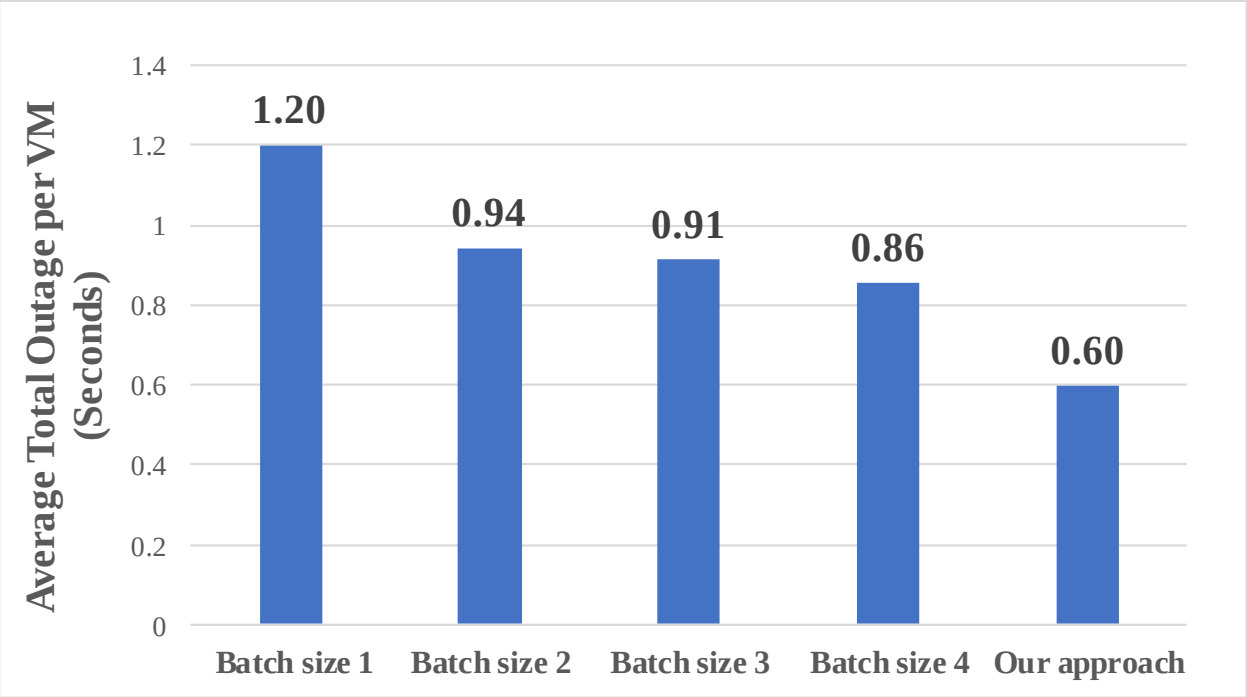

a) For evaluation scenario (a)

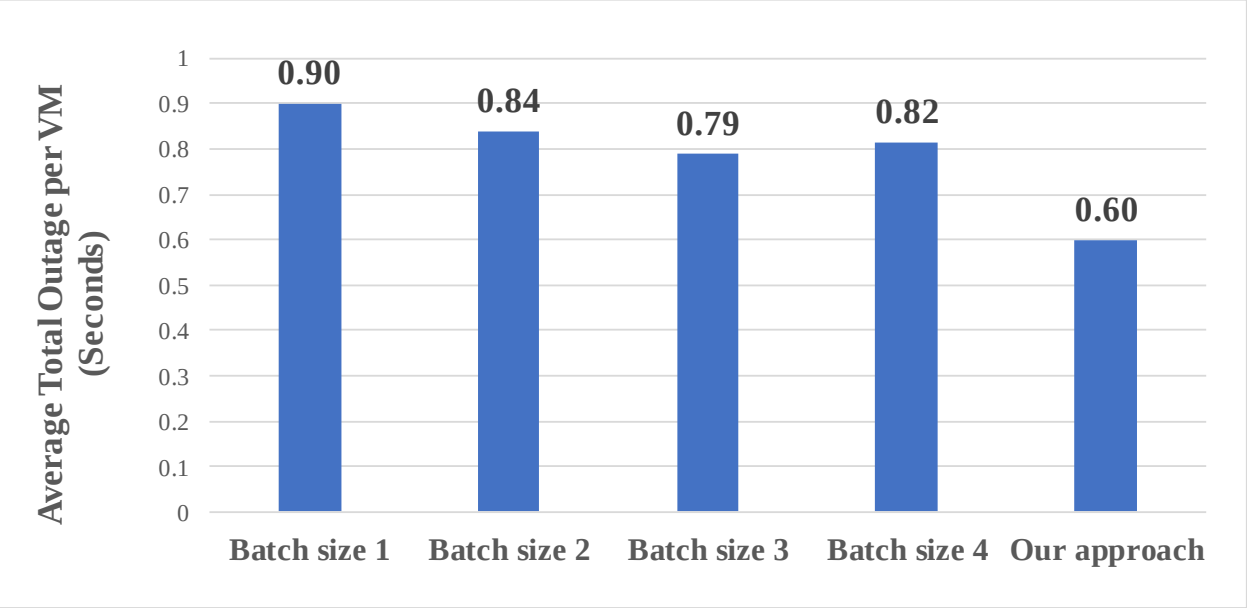

b) For evaluation scenario (b)

Fig. 11. Comparison of the outage for each VM during the upgrade

As it was expected, the results reported Table II and Table III show that by increasing the batch size in the rolling upgrade method, the duration of SLA violations per tenant decreases, as well as the duration of the upgrade. However, the number of impacted VMs, per SLA violation, increases, which causes increase in the applicable quadratic proportional penalties. The applicable penalties for our approach are less than for the rolling upgrade method, as our approach ensures that the number of impacted VMs per impact is no more than 1 VM per tenant, and it also prioritizes the upgrade of not in-use nodes which reduces the number of VM migrations during the upgrade.

Note that when the tenants are scaled out to their maximum number of VMs, the upgrade process will be paused in our approach until scaling in happens. Whereas, the rolling upgrade method will continue regardless of the state of the system. This will cause more SLA violations and increase in the applicable penalties, compared to the scenarios considered in our evaluation.

Overall, the evaluation demonstrates that our approach works better with respect to the SLA constraints of availability and elasticity, compared to the rolling upgrade method with fixed batch sizes.

### C. *Threats to validity*

The following internal and external threats can affect the validity of our experimental validation.

*Internal validity threats:* The first threat to validity is that we conducted our evaluation by calculating the measurements for a cluster of ten compute nodes considering the measurements obtained from a real deployment of six compute nodes. Experimental results from a real deployment of ten compute nodes might differ from the results calculated in our evaluation.

| Method /approach | Total duration of upgrade (seconds) | Number of SLA violations per tenant | | Number of impacted VMs per tenant in each SLA violation | | Average total duration of SLA violation per tenant (seconds) | Applicable quadratic proportional penalty per tenant |
|---|---|---|---|---|---|---|---|
| | | Min | Max | Min | Max | | |
| **Batch Size 1** | 548.00 | 2 | 6 | 1 | 1 | 2.25 | 2.25 q’ |
| **Batch Size 2** | 287.14 | 2 | 4 | 1 | 2 | 1.69 | 2.98 q’ |
| **Batch Size 3** | 226.43 | 1 | 3 | 1 | 2 | 1.35 | 3.54 q’ |
| **Batch Size 4** | 175.57 | 1 | 2 | 1 | 2 | 1.24 | 3.56 q’ |
| **Our Approach** | 192.69 | 1 | 3 | 1 | 1 | 1.35 | 1.35 q’ |

TABLE II. SLA VIOLATION RELATED MEASUREMENT RESULTS FOR ALL POSSIBLE BATCH SELECTIONS FOR EVALUATION SCENARIO (A)

| Method /approach | Total duration of upgrade (seconds) | Number of SLA violations per tenant | | Number of impacted VMs per tenant in each SLA violation | | Average total duration of SLA violation per tenant (seconds) | Applicable quadratic proportional penalty per tenant |
|---|---|---|---|---|---|---|---|
| | | Min | Max | Min | Max | | |
| **Batch Size 1** | 582.50 | 5 | 8 | 1 | 1 | 3.38 | 3.38 q’ |
| **Batch Size 2** | 300.29 | 3 | 4 | 1 | 2 | 2.20 | 4.86 q’ |
| **Batch Size 3** | 226.43 | 2 | 3 | 1 | 3 | 1.53 | 6.55 q’ |
| **Batch Size 4** | 183.69 | 2 | 3 | 1 | 3 | 1.50 | 7.3 q’ |
| **Our Approach** | 361.92 | 3 | 5 | 1 | 1 | 2.25 | 2.25 q’ |

TABLE III. SLA VIOLATION RELATED MEASUREMENT RESULTS FOR ALL POSSIBLE BATCH SELECTIONS FOR EVALUATION SCENARIO (B)

The second threat relates to the size of the cluster and the number of tenants. The results of our evaluation may vary considering a cluster with larger number of compute nodes and larger number of tenants. In addition, we used a virtual cluster in our experiments. One needs to consider a physical cluster of compute nodes and consider performing further experiments. The third threat relates to the number of experiments we performed. We ran ten sets of experiments for upgrading the cluster of six nodes and considered the average results. More experiments would be required to confirm our experiments' results. As for the fourth threat, the size of the VMs can impact the outage during live migration and subsequently the results of the evaluation. In our experiments, we used VMs with the tiny flavor. Further experiments may explore the impact of using different VM flavors on the availability measurements and the possible SLA violations.

*External validity threats*: our experiment was conducted in the context of the OpenStack cloud platform. Before generalizing the results, one needs to consider upgrades in different cloud platforms.

Nevertheless, the experiments conducted demonstrate that our approach works well with respect to the SLA constraints of availability and elasticity. The principles behind our approach and framework targeted such a property.

## VI. RELATED WORK

There are different upgrade methods proposed for maintaining HA during the upgrade of cluster based systems. However, none of these methods alone is sufficient to overcome all the challenges faced when upgrading an IaaS cloud system. These methods were designed for cluster-based HA systems and they address in isolation from one another the different challenges of upgrading such systems. To upgrade the IaaS cloud systems, these methods can be used in an upgrade orchestration framework which handles the cloud specific aspects of upgrade (e.g. dynamicity of the system).

In [9] three methods have been proposed for upgrading cluster based giant-scale systems, "fast reboot", "rolling upgrade", and "big flip". Fast reboot is proposed as the simplest upgrade method by quickly rebooting the entire system simultaneously into the new versions, however it cannot satisfy service continuity and HA of the applications running on the system. To maintain HA during the upgrade when there is no incompatibility between the versions of the nodes rolling upgrade is recommended. The nodes are upgraded one at a time like a wave rolling through the cluster. Although rolling upgrade is also introduced in [29] as one of the industry best practice, it is also criticized in [13][14][29] as it may introduce incompatibilities (referred as mixed-version inconsistencies) during the upgrade. Moreover, applying the rolling upgrade to a large system may take very long time. In addition, the rolling upgrade has to be applied separately to each different kinds of IaaS resources. This adds to the duration of the upgrade. In our approach, to minimize the duration of the upgrade, we identify the resources that can be upgraded simultaneously (while respecting dependencies and SLA constraints) and apply the rolling upgrade with dynamic batch sizes.

In the presence of incompatibilities, [9] recommends the use of the big flip method, which overcomes this challenge by upgrading one half of the system first and then flipping from the old version to the new one to prevent running two different versions at the same time [9]. Note that big flip is referred to as split mode in our and some other papers [10]. Although this method is powerful to avoid incompatibilities, it reduces the capacity of the system to its half during the upgrade, which is an issue if there is not enough redundancy in the system. To improve this upgrade method, delayed switch is proposed in [11] for upgrading cloud systems, where first the nodes are upgraded one at a time and remain deactivated after the upgrade to avoid incompatibility. When half of the system is upgraded, a switch is performed by deactivating the remaining old nodes and reactivating all the upgraded ones. Then, the remaining old nodes are upgraded simultaneously [11]. Another solution proposed for avoiding backward incompatibilities during upgrades is to use explicit embedded at development time versioning of the software [14]. However, in [14] this solution is applied to a limited set of resources, i.e. which modifies persistent data structures. It is not applicable to upgrade of different kinds of IaaS resources.

To address backward incompatibility other techniques have also been proposed, which are similar to big flip. In [13][15][16], the Imago system (also referred to as parallel universe) is presented to perform online upgrades. In this method, an entirely new system is created with the new version of the software, while the old system continues to run. Similar to split mode/big flip first, persistent data is transferred from the old system to the new one to be able to test the new system before switching over. Once the new system is sufficiently tested, the traffic is redirected to the new system [13]. This expensive solution doubles the amount of resources used during the upgrade. To minimize the amount of additional resources used during the upgrade, in our approach instead of bringing up a complete IaaS system as a parallel universe, we use this method locally to upgrade the storage/controller resources supporting VM operations.

Despite the challenge of incompatibility associated with rolling upgrade, this method is still widely used by cloud providers. Windows Azure storage use rolling upgrade to upgrade the storage system [12] by upgrading one upgrade domain (i.e. a set of evenly distributed servers and replicated storages) at a time in a rolling manner. To maintain the availability of the system during the upgrade, it is ensured to have enough storage replicas in the remaining upgrade domains [12]. Rolling upgrade is also used by Amazon Web Services (AWS) to update or replace Amazon Elastic Beanstalk (PaaS) [30], or Amazon EC2 (IaaS) [7] instances. In Amazon EC2, the rolling upgrade is applied to instances of autoscaling groups in which the batch size can be predefined. To avoid interference between upgrade and autoscaling, it is recommended to suspend autoscaling during the upgrade [7]. Disabling the autoscaling feature during upgrades, disables one of the most important features of the cloud. In our work, instead of disabling the autoscaling feature, we make it regulate the upgrade process, i.e. suspending the upgrade process or increasing the batch size depending on the resources available for upgrade.

Rolling upgrade is also used in [31] to upgrade the VM instances. In this work, the optimization problem of rolling upgrade with the multi objectives of minimizing the completion time, the cost and the expected loss of service instances (i.e. VMs) is investigated. They formalized the rolling upgrade considering different iterations of upgrade with a fixed batch size (or granularity in their terms) defined by operator. Considering potential failures during upgrades, the number of successful upgrades may be less than the predefined batch size, resulting in a longer completion time [31]. In contrast to our work, [31] does not consider changes in the number of VM instances during the upgrade due to autoscaling. It does not address the challenge of interferences between autoscaling and the upgrade process.

In [32] rolling upgrade is used for the reconfiguration of cluster membership using quorums. Subset of the servers, which have the same information replicated on them, are organized into a quorum. Any member of the quorum can become the candidate leader to initiate a configuration change. The proposed configuration with the largest ballot is selected as the target configuration. In each iteration of the upgrade, a predefined batch of servers is upgraded simultaneously. This approach is suitable for upgrading distributed state-full services (e.g. database service) except for the distributed locking service [32]. In this paper the dynamicity of the system due to autoscaling is not considered.

In [33] state aware instances are suggested to address the incompatibility issues in the rolling upgrade. The instances are upgraded from the old version to the new one using rolling upgrade. However, only instances with the old version are active until a point where switchover is performed to the new version while deactivating the old version. This method is similar to the delayed switch method, with the difference in the switching point. The switching point has to be determined based on the availability and scalability requirements of the system and its impact on the availability and the capacity of the system. Although this paper quantifies the risk associated with the version switching during the rolling upgrade, it considers neither the possible interference of the upgrade process with the autoscaling feature, nor does the upgrade of different kinds of IaaS resources.

In [34], an approach is proposed for controlling the progress of the rolling upgrade based on failures, which is referred to as "robust rolling upgrade in the cloud ($R_2C$)". In this paper, which is an extension of [33], the rolling upgrade controller controls the progress of the upgrade based on inputs from an error detection mechanism. Based on the type of the failure during the upgrade (e.g. platform/infrastructure failures and operation failures), the rolling upgrade controller decides whether to replace the problematic instance or to suspend the upgrade process if the errors impact the process of the upgrade. Similar to [33], since they replace the failed resource with the old version of the instance, in each iteration during the upgrade, the number of upgraded resources can increase or decrease. This paper provides a prediction model for the expected completion time of the rolling upgrade based on the probability of the different failures and using the different batch sizes for different runs of the upgrade. Note that in this paper, the batch size is fixed during the upgrade, and it is set by the administrator. In addition, the administrator also selects the switching point to the new version. Since the batch size is not adjusted at runtime to the current state of the system with respect to the SLAs of the tenants, autoscaling may interfere with the upgrade process. In contrast, our approach regulates the upgrade process based on considerations for potential scaling out requests to minimize such interference. More importantly, [34] targets only the upgrade of VM instances, while our approach handles the upgrade of different IaaS resources.

Although all the above-mentioned upgrade methods address the problem of maintaining availability and in some cases the challenge of incompatibilities, they do not address all the challenges the upgrade of cloud systems poses. In particular they do not address the different kinds of dependencies and the dynamicity of the cloud. In contrast, in this paper we propose an upgrade management framework for handling all these aspects of upgrades of the IaaS cloud system in an integrated manner.

## VII. Conclusion

In this paper, we proposed a novel approach for the management of upgrade of IaaS cloud systems under SLA constraints for availability and elasticity. In this method, we tackled in an integrated manner the challenges posed by the dependencies and the possible incompatibilities along the dependencies, by the potential upgrade failures, by the dynamicity of the IaaS cloud system, and by the amount of extra resources used during upgrade.

In the proposed approach, an upgrade is initiated by an upgrade request which is composed of change sets requested for example by a system administrator indicating the desired changes in IaaS cloud system. In addition to the initial change sets, the proposed method allows for new upgrade requests to be issued continuously during the upgrade process. The upgrade actions required to upgrade each IaaS resource, the upgrade method appropriate for each subset of resources, and the batch of resources to upgrade in each iteration are determined automatically and applied in an iterative manner. Since in each iteration, the batch of resources to upgrade is selected according to the current state of the system with respect to the dependencies and the SLA constraints, the inference between autoscaling and the upgrade process is mitigated. Furthermore, since the upgrade process is regulated based on the current state of the system, cloud providers can perform the upgrades gradually according to the state of the system, and they do not need to designate a maintenance window for performing the upgrades. In the proposed method, in case of upgrade failures, localized retry and undo operations are also issued automatically according to the failures and undo/retry thresholds indicated by the administrator. This feature has the capability to undo a failed change set, while the upgrade proceeds with other change sets.

We have implemented a proof of concept demonstrating the feasibility of the proposed upgrade management framework for the upgrade of IaaS compute and its application in an OpenStack cluster. We conducted experiments to show how our approach works with respect to the SLA constrains of availability and elasticity. The experiments demonstrate that our approach does not introduce any outage at the application level for the tenants whose VMs are configured as a HA cluster at the application level. Moreover the results indicate that our approach reduces the SLA violations during the upgrade, compared to the rolling upgrade with fixed batch sizes.

## Acknowledgment

This work has been partially supported by the Natural Sciences and Engineering Research Council of Canada (NSERC) and Ericsson.

## APPENDIX

| Symbols | Description | Symbols | Description |
|---|---|---|---|
| *K, K'* | Host capacity in terms of VMs (before and after a hypervisor upgrade) | $M_{network}$ | Set of hosts dedicated to networking services |
| $N_i$ | Number of tenants in iteration *i* | $M_{controller}$ | Set of hosts dedicated to controller services |
| $min_n$ | Minimum number of VMs for tenant *n* | $M_{computeForOldVM}$ | Set of compute hosts capable of hosting VMs of the old version |
| $max_n$ | Maximum number of VMs for tenant *n* | $M_{computeForNewVM}$ | Set of compute hosts capable of hosting VMs of the new version |
| $c_n$ | Cooldown time for tenant *n* | $M_{usedCompute}$ | Set of in-use compute hosts |
| $s_n$ | Scaling adjustment in terms of VMs per cooldown time for tenant n | $M_{usedComputeForOldVM}$ | Set of in-use compute hosts with VMs of the old version |
| $S_i$ | Maximum scaling adjustement requests per tenant that may occur during iteration *i* | $M_{usedComputeForNewVM}$ | Set of in-use compute hosts with VMs of the new version |
| $T_i$ | The time required to upgrade and to recover from potential failures of the batch of iteration *i* | $ScalingResv_{forOldVM}$ | Number of compute hosts reserved for scaling of VMs of the old version |
| *F* | The number of compute host failures to be tolerated during an iteration | $ScalingResv_{forNewVM}$ | Number of compute hosts reserved for scaling of VMs of the new version |
| $A_i$ | Number of tenants who might scale out on hosts compatible with the old VM version in iteration *i* | $FailoverResev_{forOldVM}$ | Number of compute hosts reserved for failover of VMs of the old version |
| $Z_i$ | The maximum number of compute hosts that can be taken out of service in iteration *i* | $FailoverResev_{forNewVM}$ | Number of compute hosts reserved for failover of VMs of the new version |
| $V_i$ | The total number of VMs to be upgraded in iteration *i* | $MinHostReqConf_{oldStorage}$ | Minimum required number of storage hosts for the old configuration of the virtual storage |
| $W_{ij}$ | The batch size in terms of VMs where each VM belongs to a different anti-affinity group in the main iteration *i* and sub-iteration *j* | $MinHostReqConf_{newStorage}$ | Minimum required number of storage hosts for the new configuration of the virtual storage |
| $M_{storage}$ | Set of hosts eligible to participate in the creation of virtual storage (storage hosts) | $MinHostReqCap_{oldStorage}$ | Minimum required number of storage hosts for data of VMs of the old version |
| $M_{compute}$ | Set of hosts eligible to provide compute services (compute hosts) | $MinHostReqCap_{newStorage}$ | Minimum required number of storage hosts for data of VMs of the new version |
| $G_{batch}$ | The initial batch, the resource groups that can be upgraded in the current iteration without violating any of their dependencies | $G_{Fbatch}$ | The final batch, the selected resource groups for the upgrade after considering SLA constraints |

TABLE A. DEFINITION FOR THE PARAMETERS USED IN THE PROPOSED APPROACH